\documentclass[11pt]{article}
\usepackage{graphicx,epsfig,color}
\usepackage{cite}
\renewcommand{\theequation}{\arabic{section}.\arabic{equation}}
\renewcommand{\thesection}{\arabic{section}.}

\renewcommand{\thefootnote}{\fnsymbol{footnote}}
\newcommand{\bc}{\begin{center}}
\newcommand{\ec}{\end{center}}
\newcommand{\be}{\begin{equation}}
\newcommand{\ee}{\end{equation}}
\newcommand{\bea}{\begin{eqnarray}}
\newcommand{\eea}{\end{eqnarray}}
\newcommand{\ba}{\begin{array}}
\newcommand{\ea}{\end{array}}
\newcommand{\lb}{\label}
\newcommand{\rf}{\ref}
\newcommand{\bfg}{\begin{figure}[htbp]}
\newcommand{\efg}{\end{figure}}

\newcommand{\prd}{Phys. Rev. D }
\newcommand{\np}{Nucl. Phys. }
\newcommand{\npb}{Nucl. Phys. B }
\newcommand{\prl}{Phys. Rev. Lett. }
\newcommand{\prp}{Phys. Rep. }

\newcommand{\pl}{Phys. Lett. }

\newcommand{\ptp}{Prog. Theor. Phys. }
\newcommand{\ptep}{Prog. Theor. Exp. Phys. }

\begin{document} 

\vspace*{1. cm}
\bc
{\Large {\boldmath \textbf{Tetraquark and two-meson states at 
large ${N_{\mathrm{c}}}$}}}
\\
\vspace{1 cm}
{Wolfgang Lucha$^a$, Dmitri Melikhov$^{a,b,c}$, Hagop Sazdjian$^d$}\\
\vspace{0.5 cm}
{\small
{$^a$Institute for High Energy Physics, Austrian Academy of 
Sciences, Nikolsdorfergasse 18,\\ 
A-1050 Vienna, Austria\\
$^b$D.~V.~Skobeltsyn Institute of Nuclear Physics,
M.~V.~Lomonosov Moscow State University,\\ 
119991, Moscow, Russia\\
$^c$Faculty of Physics, University of Vienna, Boltzmanngasse 5, 
A-1090 Vienna, Austria\\ 
$^d$IPNO, Universit\'e Paris-Sud, CNRS-IN2P3, Universit\'e Paris-Saclay, 
91405 Orsay, France}\\
E-mail:  wolfgang.lucha@oeaw.ac.at , dmitri\_melikhov@gmx.de , 
sazdjian@ipno.in2p3.fr
}
\ec
\par
\vspace{0.5 cm}

\bc
{\large Abstract}
\ec
Considering four-point correlation functions of color-singlet 
quark bilinears, we investigate, in the large-$N_{\mathrm{c}}^{}$ limit
of QCD, the subleading diagrams that involve, in the $s$-channel
of meson-meson scattering amplitudes, two-quark--two-antiquark 
intermediate states. The latter contribute, together with gluon 
exchanges, to the formation, at the hadronic level, of two-meson and 
tetraquark intermediate states. It is shown that the two-meson
contributions, which are predictable, in general, from leading-order
$N_{\mathrm{c}}^{}$-behaviors, consistently satisfy the constraints 
resulting from the $1/N_{\mathrm{c}}^{}$ expansion procedure and thus 
provide a firm basis for the extraction of tetraquark properties from 
$N_{\mathrm{c}}^{}$-subleading diagrams. We find that, in general, 
tetraquarks, if they exist in compact form, should have narrow decay 
widths, of the order of $N_{\mathrm{c}}^{-2}$. For the particular case 
of exotic tetraquarks, involving four different quark flavors, two 
different types of tetraquark are needed, each having a preferred decay 
channel, to satisfy the consistency constraints.
\par
\vspace{0.25 cm}
{\small
}
\par 
\vspace{0.75 cm}
\renewcommand{\thefootnote}{\arabic{footnote}}

\section{Introduction} \lb{s1}
\setcounter{equation}{0}

The existence of tetraquarks as tightly bound states of QCD 
\cite{Jaffe:1976ih,Jaffe:2008zz}, also called compact tetra\-quarks, 
is still a matter of theoretical debate. The problem is related to 
the issue as to whether the interquark confining forces may produce 
bound states made essentially of a pair of quarks and a pair of 
antiquarks, just as they produce ordinary meson states made of a 
quark and an antiquark. 
A questioning may arise from the fact that interpolating color-neutral 
four-quark local operators that would create tetraquark states can 
be decomposed by Fierz rearrangements into combinations of products 
of color-neutral quark bilinears \cite{Nielsen:2009uh}, which are 
rather suggestive of the creation of pairs of free ordinary meson 
states. In the absence of an exact resolution of the four-body 
problem in the presence of confining forces, simplified models, 
based on the diquark and antidiquark associations, have been proposed, 
in which the existence of attractive forces, favoring the formation 
of tetraquark bound states, is more transparent  
\cite{Jaffe:2003sg,Maiani:2004vq,Ebert:2007rn,Czarnecki:2017vco,
Lebed:2017min}. Lattice calculations do not yet bring firm conclusions, 
the results often depending on the flavor of the heavy quarks that are 
considered \cite{Ikeda:2016zwx,Francis:2016hui,Bicudo:2016ooe,
Cheung:2017tnt,Hughes:2017xie}.   
\par
On the other hand, it is generally admitted that in 't Hooft's 
large-$N_{\mathrm{c}}^{}$ limit of QCD \cite{'tHooft:1974hx}, with the 
coupling constant $g$ scaling as $N_{\mathrm{c}}^{-1/2}$ and with the 
quark fields belonging to the fundamental representation of the 
color gauge group $\mathrm{SU}(N_{\mathrm{c}}^{})$, the theory catches 
the main properties of confinement, while being liberated from secondary 
screening phenomena, such as quark pair creation or inelasticity effects. 
Applying this approach to color-neutral quark bilinear operators, Witten 
has shown that in this limit the related QCD correlation functions are 
saturated by noninteracting meson states \cite{Witten:1979kh}. 
Generalizing the application to color-neutral quark quadrilinear operators, 
Coleman then showed that the correlation functions of the latter are 
dominated, in the large-$N_{\mathrm{c}}^{}$ limit, by free ordinary meson 
states \cite{Coleman:1985}.
\par
For a long time, the latter result has been considered as an indication
for the nonexistence of tetraquarks as bound states surviving the
large-$N_{\mathrm{c}}^{}$ limit. Recently, however, Weinberg has 
reexamined the question by noticing the fact that the subleading nature 
of the interaction part of the quark quadrilinear operators is not a 
proof of the nonexistence of tetraquarks, but rather might be a constraint
on their decay widths. Considering a class of candidate operators,
he showed that if tetraquarks exist in the large-$N_{\mathrm{c}}^{}$ limit 
as bound states with finite masses, then they should have narrow widths,
of the order of $N_{\mathrm{c}}^{-1}$, like those of the ordinary mesons, 
and would thus be observable \cite{Weinberg:2013cfa}. In complement to the
latter result, Knecht and Peris have stressed that depending on their 
flavor content, tetraquarks might even be narrower in some cases, 
having widths of the order of $N_{\mathrm{c}}^{-2}$ \cite{Knecht:2013yqa}. 
Cohen and Lebed, studying the analyticity properties of meson-meson 
scattering amplitudes, have reported that in the case of exotic 
tetraquarks the decay widths should, in general, be of the order of 
$N_{\mathrm{c}}^{-2}$ or less \cite{Cohen:2014tga}. 
Ref. \cite{Maiani:2016hxw} reported the possibility of smaller widths
of the order of $N_\mathrm{c}^{-3}$. The $N_\mathrm{c}^{}$-analysis
of meson-meson scattering amplitudes, in connection with lattice
calculations, has also been presented in \cite{Guo:2013nja}.    
\par
The aim of the present paper is to investigate in a systematic way
the $N_{\mathrm{c}}^{}$-subleading diagrams where tetraquark candidates 
may occur. They are characterized by the presence of 
two-quark--two-antiquark intermediate states\footnote{Henceforth called 
for simplicity ``four-quark intermediate states''.} in the $s$-channel 
of meson-meson scattering amplitudes. 
Here, however, an additional complication arises with respect 
to the usual cases of $N_{\mathrm{c}}^{}$-leading diagrams: four-quark 
intermediate states also signal the presence of two interacting meson 
states at the hadronic level. Since the $N_{\mathrm{c}}^{}$-behavior of 
three-meson and four-meson vertices can be determined, in general, from 
simpler diagrams, the subleading two-meson contributions are then 
completely predicted and thus should satisfy consistency checks within 
the above analysis. The latter is a crucial test for the validity of
the $1/N_{\mathrm{c}}^{}$-expansion method in QCD. It is once that these
contributions are evaluated and tested that one may safely extract
the properties of tetraquarks from the $N_{\mathrm{c}}$-subleading 
diagrams.
\par
Concerning the four-quark intermediate states, their presence should
be determined with the aid of the Landau equations 
\cite{Landau:1959fi,Itzykson:1980rh}, which provide unambiguous 
criteria for their existence. It is also understood that each QCD 
diagram with four-quark intermediate states is accompanied 
by similar diagrams with insertions of any number of gluon lines neither 
changing its topology, nor its $N_{\mathrm{c}}$-behavior; it is the 
infinite sum of such diagrams that produces the hadronic-state 
singularities, as tetraquark poles or two-meson cuts.    
\par
An important ingredient in the present approach is provided by the
consideration, for a given set of quark flavors, of all possible
meson-meson scattering channels which may produce four-quark
singularities; in this way, one obtains the maximum number of 
constraints on the properties of the tetraquark candidates, which 
often are not apparent within a single channel. 
We also emphasize that no hypothesis is done about the internal 
structure of the tetraquark states with respect to the possible 
combinations of quark fields. In some cases, the existing constraints
are strong enough to suggest the most favorable structures.
\par
Our main results can be summarized as follows. First, the two-meson
contributions, which emerge through effective meson one-loop diagrams,
satisfy all the consistency checks coming from the 
$N_{\mathrm{c}}$-behaviors of three- and four-meson vertices. 
This confirms the validity of the 
perturbative $1/N_{\mathrm{c}}^{}$ expansion to the next-to-leading-order 
diagrams. Second, tetraquarks, if they exist, should have, in general,
narrow decay widths, of the order of $N_{\mathrm{c}}^{-2}$, much smaller 
than those of ordinary mesons, which are of the order of 
$N_{\mathrm{c}}^{-1}$. Third, in 
the case of exotic sectors, involving four different quark flavors, two 
different tetraquarks, each having a preferred decay channel, are needed 
to fulfill the consistency conditions. The internal structure of these
tetraquarks, expressed as a quark quadrilinear, would be of the form
of a product of two color-singlet bilinears. Part of the results above 
has been presented in \cite{Lucha:2017mof}.   
\par
The generality of the tetraquark width estimate of being of the order
of $N_{\mathrm{c}}^{-2}$, obtained in the present paper, in 
contradistinction with the weaker estimate of Ref. \cite{Weinberg:2013cfa}, 
$O(N_{\mathrm{c}}^{-1})$, stems from the fact that color-planar
diagrams, with one external quark loop and without internal quark loops,
do not have $s$-channel four-quark singularities. This eliminates 
the potential presence of tetraquark intermediate states in such types of 
diagram and consequently reduces the magnitude of the tetraquark--two-meson
transition amplitudes. This feature has been overlooked in Ref. 
\cite{Weinberg:2013cfa}.
\par
We concentrate, in the following (Secs. 2 and 3), on the 
cases of exotic and cryptoexotic channels, corresponding to four and 
three different quark flavors, respectively. The case of two flavors 
can be treated in a similar way as for three and is briefly sketched. 
Details of the calculations related to the role of the Landau equations 
are presented in the appendix.
\par
For recent reviews on the experimental properties of tetraquark 
candidates and their theoretical interpretations, we refer 
the reader to Refs. \cite{Cohen:2014vta,Esposito:2014rxa,Olsen:2014qna,
Chen:2016qju,Hosaka:2016pey,Esposito:2016noz,Karliner:2016joc,
Lebed:2016hpi,Guo:2017jvc,Ali:2017jda,Olsen:2017bmm}.
\par  
%\newpage 

\section{Exotic channels} \lb{s2}
\setcounter{equation}{0}

Our analysis is based on the study of four-point correlation functions 
of color-singlet meson sources or currents $j$ of the type 
$\langle jjj^{\dagger}j^{\dagger}\rangle$, where the $j$'s are specified 
by their quark flavor content. We define
\be \lb{2e1}
j_{ab}^{}=\overline q_a^{}\hat Oq_b^{},
\ee
where $a$ and $b$ are flavor indices and $\hat O$ is a combination of
Dirac matrices. We do not focus in this work on the detailed spin 
and parity structure of mesons and tetraquarks, which does not play 
a fundamental role in the subsequent analyses, and hence shall omit
all Lorentz structures. 
An ordinary meson, having the flavor content of antiquark $a$ and of 
quark $b$ will be designated by $M_{ab}^{}$; its coupling 
to the current $j_{ab}^{}$ is designated by $f_{M_{ab}^{}}^{}$:
\be \lb{2e2}
\langle 0|j_{ab}^{}|M_{ab}^{}\rangle = f_{M_{ab}^{}}^{}
\ =\ O(N_{\mathrm{c}}^{1/2}),
\ee
where we have also indicated its large-$N_{\mathrm{c}}^{}$ behavior
\cite{Witten:1979kh}.  
\par 
We concentrate in this section on fully exotic channels, in which 
four different quark flavors are present, designated by the labels
$a,b,c,d$. We consider the four-point correlation functions,
classified in the $s$-channels as ``direct'' I and II and 
``recombination'', respectively,
\be
\lb{2e3}
\Gamma_{\mathrm{I}}^{\mathrm{dir}}=\langle j_{ab}^{}j_{cd}^{}
j_{cd}^{\dagger}j_{ab}^{\dagger}\rangle,\ \ \ \  
\Gamma_{\mathrm{II}}^{\mathrm{dir}}=\langle j_{ad}^{}j_{cb}^{}
j_{cb}^{\dagger}j_{ad}^{\dagger}\rangle,\ \ \ \  
\Gamma^{\mathrm{rec}}=\langle j_{ab}^{}j_{cd}^{}j_{cb}^{\dagger}
j_{ad}^{\dagger}\rangle,\ \ \ \ \ \ a\neq b\neq c\neq d.
\ee
\par 
We first consider the direct channels. The corresponding correlation
functions have a disconnected part representing the propagation of 
two free mesons $M_{ab}^{}$ and $M_{cd}^{}$, or $M_{ad}^{}$ and 
$M_{cb}^{}$, and producing a global dependence of leading order 
$N_{\mathrm{c}}^2$, and a connected part, containing at least two gluon 
exchanges between the disconnected pieces, and having a leading-order 
behavior of $N_{\mathrm{c}}^0$ (Fig. \rf{2f1}).
\bfg 
\bc
\epsfig{file=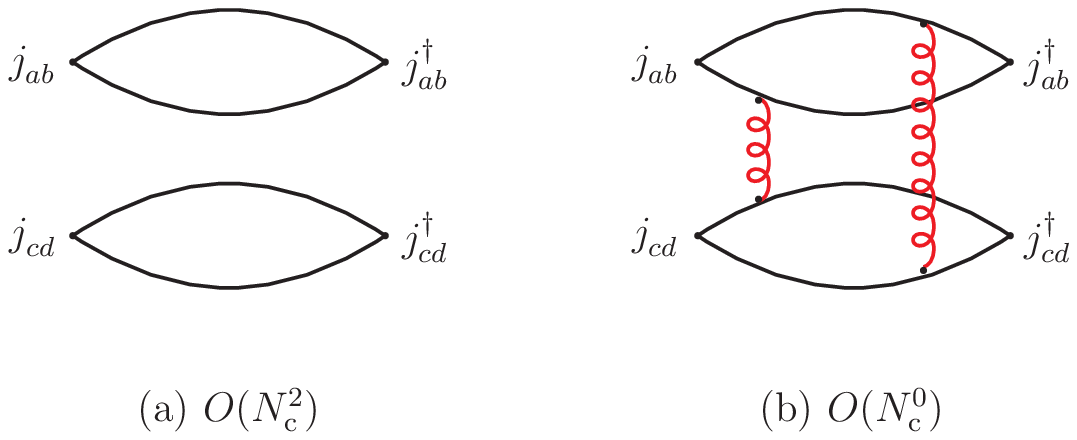,scale=0.75}
\caption{Leading- and subleading-order diagrams of the ``direct'' 
channel I of (\rf{2e3}). (a): disconnected part; (b): connected part 
(a sample diagram). Full lines represent quarks, curly lines gluons.
Similar diagrams also exist for the ``direct'' channel II.} 
\lb{2f1} 
\ec
\efg
\par
It is evident that only the connected part of the above correlation
functions can have any information about the meson-meson interaction. 
To isolate the latter, one extracts from the connected part of the 
correlation function the related scattering amplitude by factorizing 
the external four meson propagators together with the related couplings
(\rf{2e2}). These diagrams have four-quark singularities in the 
$s$-channel (cf. the appendix) and hence are saturated, at 
$N_{\mathrm{c}}$-leading order, by intermediate states composed of two 
interacting mesons and of tetraquarks, designated by $T$. 
One obtains the following leading-order behaviors for the 
two-meson scattering amplitudes and the transition amplitudes
through two-meson and tetraquark intermediate states:
\bea
\lb{2e4}
& &A(M_{ab}^{}M_{cd}^{}\rightarrow M_{ab}^{}M_{cd}^{})\ \sim\
A(M_{ad}^{}M_{cb}^{}\rightarrow M_{ad}^{}M_{cb}^{})\ =\ 
O(N_{\mathrm{c}}^{-2}),\\
\lb{2e5}
& &A(M_{ab}^{}M_{cd}^{}\rightarrow MM\rightarrow 
M_{ab}^{}M_{cd}^{})\ \sim\ 
A(M_{ad}^{}M_{cb}^{}\rightarrow MM\rightarrow 
M_{ad}^{}M_{cb}^{})\ =\ O(N_{\mathrm{c}}^{-2}),\\
\lb{2e6}
& &A(M_{ab}^{}M_{cd}^{}\rightarrow T\rightarrow M_{ab}^{}M_{cd}^{})\ 
\sim\ A(M_{ad}^{}M_{cb}^{}\rightarrow T\rightarrow M_{ad}^{}M_{cb}^{})\ 
=\ O(N_{\mathrm{c}}^{-2}).
\eea
\par
Next, we consider the recombination channel of (\rf{2e3}). Here, there 
are no disconnected diagrams and the leading-order behavior is 
$O(N_{\mathrm{c}})$ (Fig. \rf{2f2}).
\bfg 
\bc
\epsfig{file=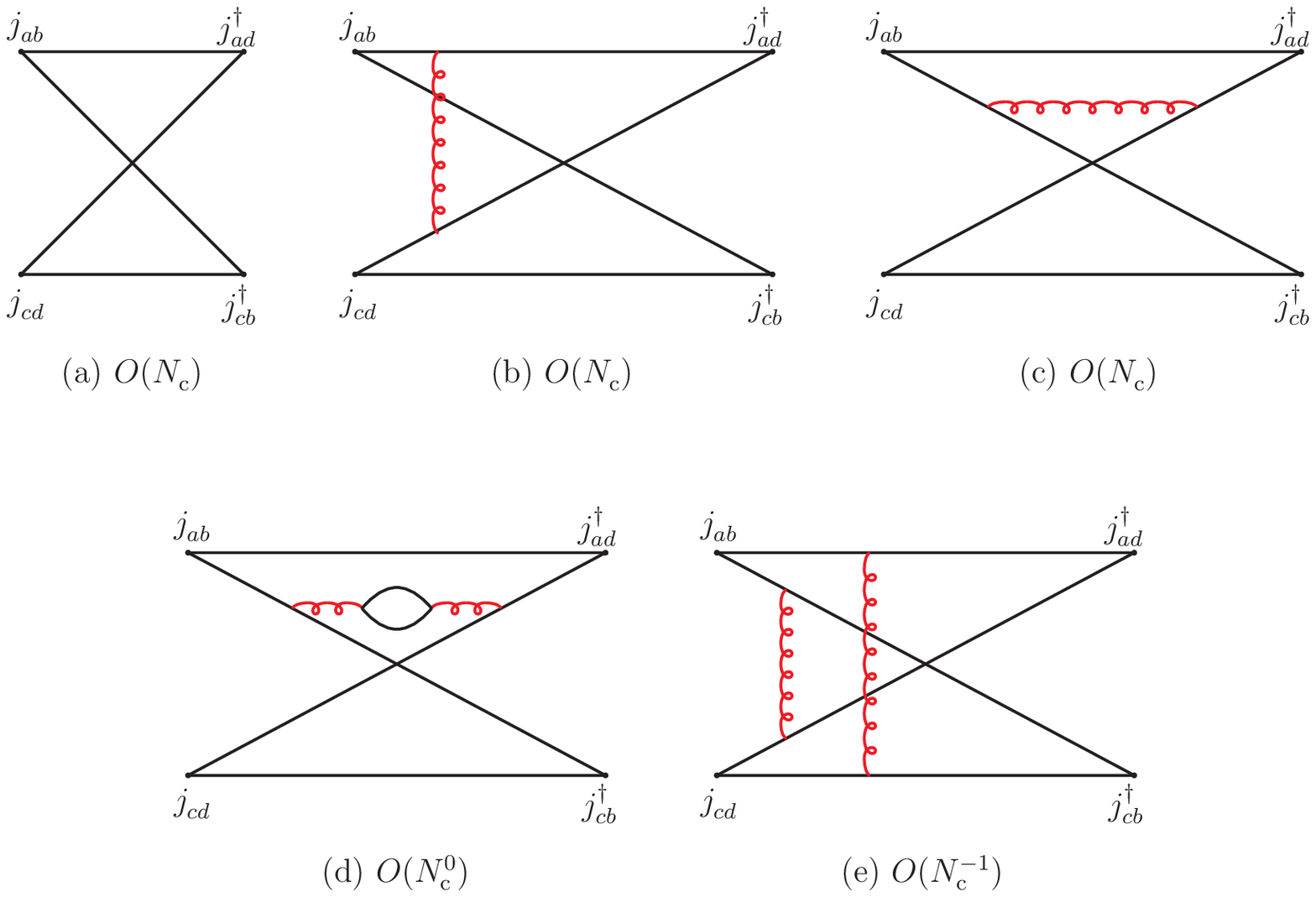,scale=0.75}
\caption{Leading- and typical subleading-order diagrams of the 
``recombination'' channel of (\rf{2e3}).}
\lb{2f2} 
\ec
\efg
\par
Using the Landau equations, one checks that diag\-ra\-ms 
\rf{2f2}(a)--\rf{2f2}(d) do not have $s$-channel singularities 
[cf. Appendix]. Their singularities arise in the $u$- and $t$-channels.
(However, diagrams \rf{2f2}(a)--\rf{2f2}(c) do not have four-quark cuts 
in any channel.) Diagram  \rf{2f2}(e) is the first diagram where 
four-quark singularities appear in the $s$-channel and hence it may 
contribute to meson-meson scattering with two-meson and tetraquark 
intermediate states. 
\par
%\begin{sloppypar}
The previous properties can also be understood in terms of the 
topological properties of the diagrams in color-space.
Diagrams \rf{2f2}(a)--\rf{2f2}(d) are color-planar, while diagram \rf{2f2}(e)
is color-nonplanar. This can be more easily seen by unfolding the diagrams
to make the color flow apparent (Fig. \rf{2f3}). The unfolded plane
corresponds now to the $(u,t)$ plane. The manifest singularities in the
color-planar diagrams correspond to those of the $u$- and $t$-chan\-nels, 
obtained with vertical and horizontal cuts, respectively. $s$-channel
singularities can be searched for by cutting the box-diagrams with
oblique and curved lines passing through the four quark propagators.
However, when the diagram is color-planar, these cuts produce, 
generally, disconnected singularities concentrated at opposite corners
and corresponding to radiative corrections of the external meson 
propagators and of the current vertices. Therefore no $s$-channel
singularities arise here. The latter may arise only when the diagrams 
are color-nonplanar, because of the specific routing of the momenta. This 
is precisely the case of diagram \rf{2f3}(e) (or its equivalent \rf{2f2}(e).) 
These properties do not depend on the number and configuration of
gluon lines, but only on the color-topology of the diagram and could
be verified on explicit examples.
%\end{sloppypar}
\par
\bfg 
\bc
\epsfig{file=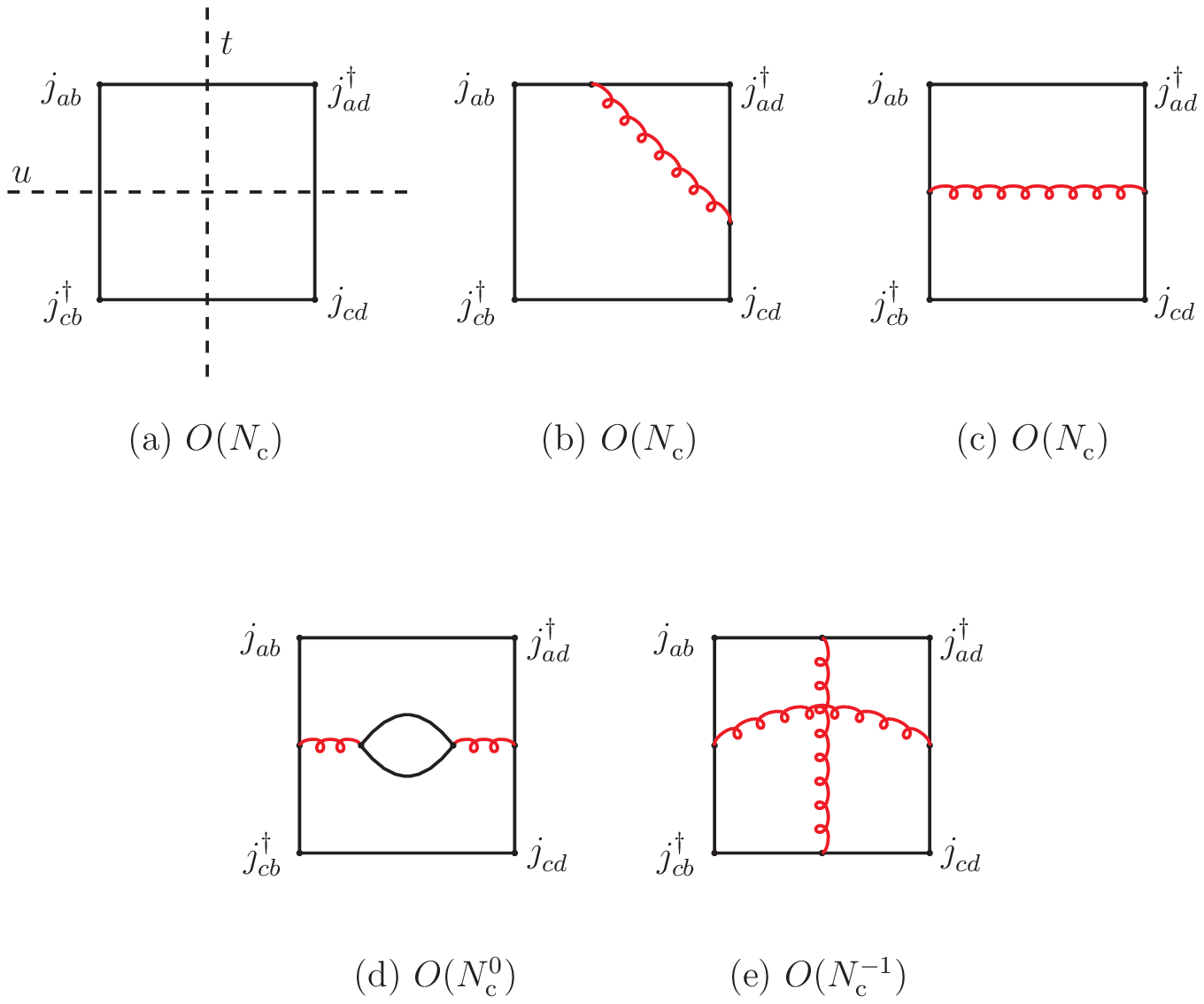,scale=0.8}
\caption{The diagrams of Fig. \rf{2f2} in unfolded form.}
\lb{2f3} 
\ec
\efg
%\begin{sloppypar}
Diagrams of the type of \rf{2f2}(d) or \rf{2f3}(d) involve, through
the internal quark loop creation, four-quark intermediate states
and might participate in the formation of tetraquark states in
the $u$-channel. However, since the internal quark loop involves
a quark and an antiquark corresponding to the same flavor, the
resulting tetraquark would belong to the class of cryptoexotic
states; the latter type of tetraquark will be explicitly considered
in Sec. 3 and hence $u$- and $t$-channel four-quark singularities
will not be analyzed here.
%\end{sloppypar}
\par
Diagrams \rf{2f2}(a)--\rf{2f2}(d) also contribute to four-meson vertex-type 
couplings, which are free of singularities.
\par
One obtains the following behaviors for the two-meson scattering 
amplitudes and the transition amplitudes through two-meson and 
tetraquark intermediate states:
\bea
\lb{2e7}
& &A(M_{ab}^{}M_{cd}^{}\rightarrow M_{ad}^{}M_{cb}^{})\ \sim\
A(M_{ad}^{}M_{cb}^{}\rightarrow M_{ab}^{}M_{cd}^{})\ =\ 
O(N_{\mathrm{c}}^{-1}),\\
\lb{2e8}
& &A(M_{ab}^{}M_{cd}^{}\rightarrow MM\rightarrow 
M_{ad}^{}M_{cb}^{})\ \sim\ 
A(M_{ad}^{}M_{cb}^{}\rightarrow MM\rightarrow 
M_{ab}^{}M_{cd}^{})\ =\ O(N_{\mathrm{c}}^{-3}),\\
\lb{2e9}
& &A(M_{ab}^{}M_{cd}^{}\rightarrow T\rightarrow M_{ad}^{}M_{cb}^{})\ 
\sim\ A(M_{ad}^{}M_{cb}^{}\rightarrow T\rightarrow M_{ab}^{}M_{cd}^{})\ 
=\ O(N_{\mathrm{c}}^{-3}).
\eea
\par
We first analyze Eqs. (\rf{2e4}) and (\rf{2e7}) in terms of effective
meson vertices.
One deduces that the four-meson vertices of the direct 
type are of order $N_{\mathrm{c}}^{-2}$, while that of the recombination 
type is of order $N_{\mathrm{c}}^{-1}$ (Figs. \rf{2f4} and \rf{2f5}):
%\newpage
\bea 
\lb{2e10}
& &g(M_{ba}^{}M_{dc}^{}M_{ab}^{}M_{cd}^{})\ \sim\
g(M_{da}^{}M_{bc}^{}M_{ad}^{}M_{cb}^{})\ =\ O(N_{\mathrm{c}}^{-2}),\\
\lb{2e11}
& &g(M_{da}^{}M_{bc}^{}M_{ab}^{}M_{cd}^{})\ =\ O(N_{\mathrm{c}}^{-1}).
\eea
\par
\bfg 
\bc
\epsfig{file=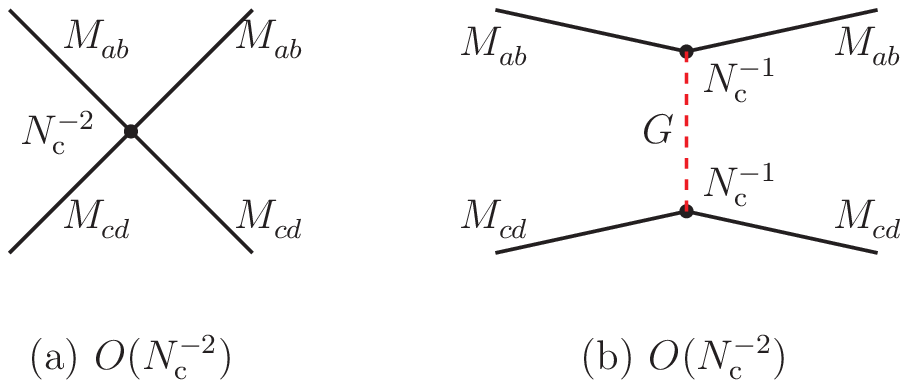,scale=0.75}
\caption{(a): Four-meson vertex in the direct channel I of (\rf{2e3});
(b): Glueball exchange in the same channel. Similar diagrams also exist 
in the direct channel II.}
\lb{2f4} 
\ec
\efg
\par
\bfg 
\bc
\epsfig{file=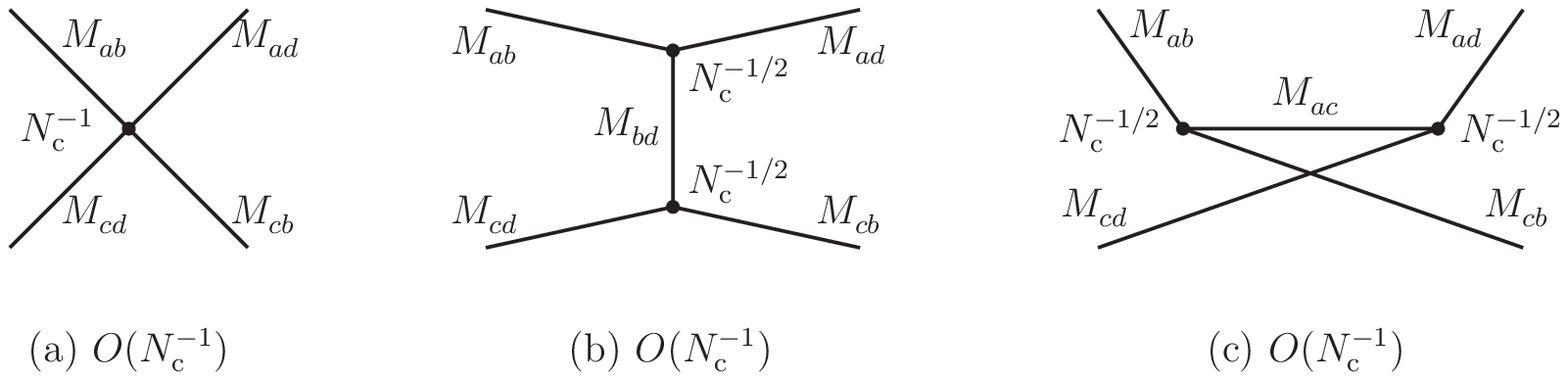,scale=0.75}
\caption{Leading-order meson diagrams in the recombination
channel; diagrams of the type of Fig. \rf{2f2}(d) contribute as 
meson loop radiative corrections or as a meson--cryptoexotic tetraquark 
mixing to diagram (c) above.}
\lb{2f5} 
\ec
\efg
\par
The difference of behavior between the two types of coupling is 
a consequence of the topological difference between the connected
recombination and direct diagrams. The former [Fig. \rf{2f2}(a)] is 
planar, the latter [Fig. \rf{2f1}(b)] is typical of OZI-suppressed 
diagrams \cite{Okubo:1963fa,Zweig:1964jf,Iizuka:1966fk}, made
connected by gluon exchanges between two disconnected pieces.
This can also be checked in theoretically founded meson effective theories.
Since our evaluations do not depend upon masses and momenta, one can
consider chiral perturbation theory \cite{Gasser:1984gg}, extended 
to $\mathrm{SU}(4)\times \mathrm{SU}(4)$. It can be verified that at the 
tree level, where four-meson couplings are, in general, of order 
$N_{\mathrm{c}}^{-1}$, the vertices of the direct channels of (\rf{2e3}) 
are absent, thus confirming the results (\rf{2e10}).
\par   
With the properties of four-meson vertices determined, one can then 
evaluate the contributions of the $s$-channel two-meson intermediate 
states in the above processes. The results are summarized in the 
diagrams of Fig. \rf{2f6}. One observes that they consistently reproduce
the behaviors expected from (\rf{2e5}) and (\rf{2e8}), corresponding
to the diagrams of Figs. \rf{2f1}(b) and \rf{2f2}(e). In particular,
going back to Fig. \rf{2f1}(b) and cutting the diagram with a vertical 
line between the gluon lines, one finds that the intermediate states
are created by the singlet operators that make the mesons $M_{ad}^{}$
and $M_{cb}^{}$, which precisely are the intermediate states of Fig. 
\rf{2f6}(a). A similar check can also be done with the other diagrams.  
\par
\bfg 
\bc
\epsfig{file=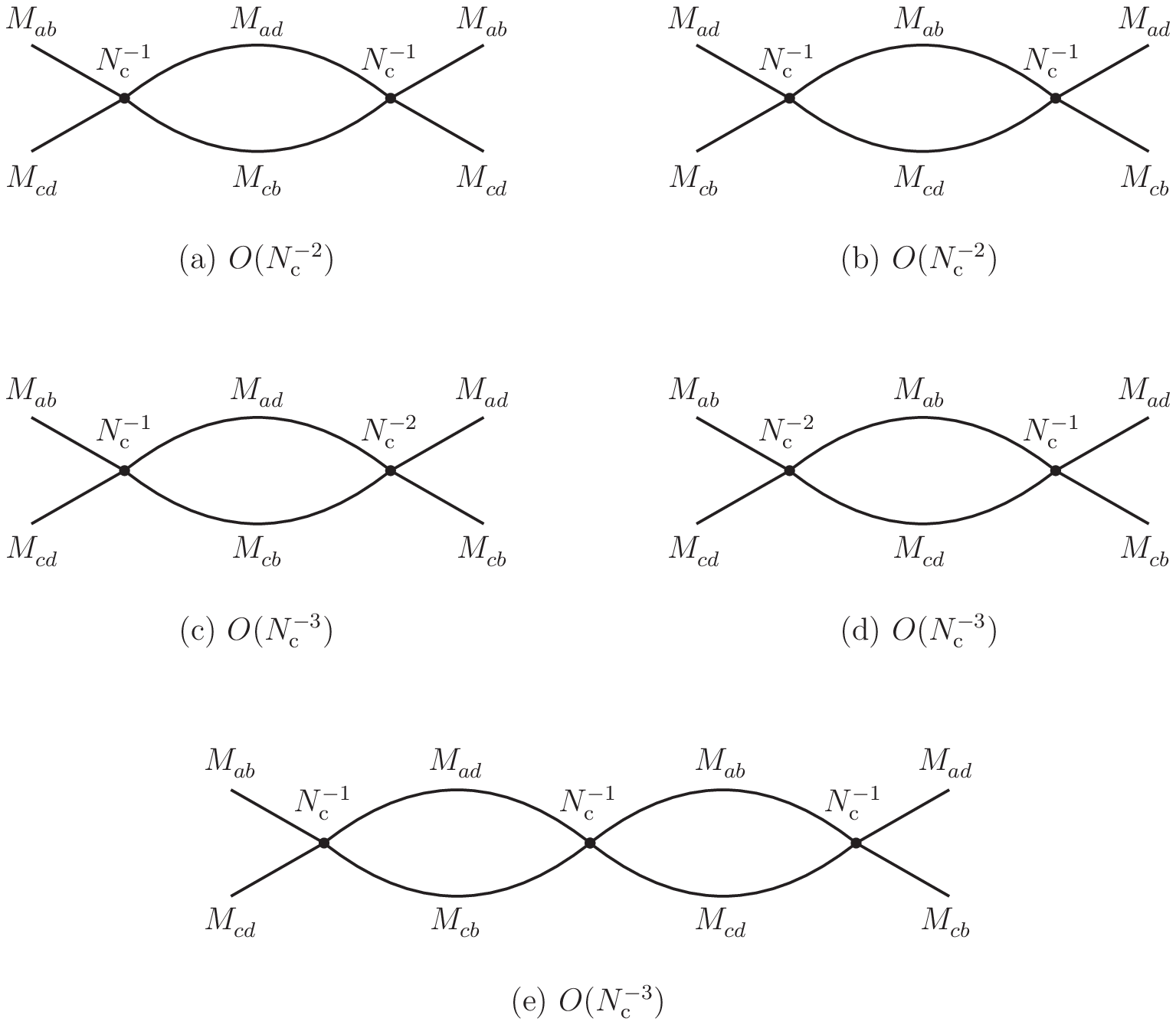,scale=0.75}
\caption{Leading-order contributions of two-meson states to the direct 
[(a) and (b)] and recombination [(c), (d) and (e)] channels.}
\lb{2f6} 
\ec
\efg 
\bfg 
\bc
\epsfig{file=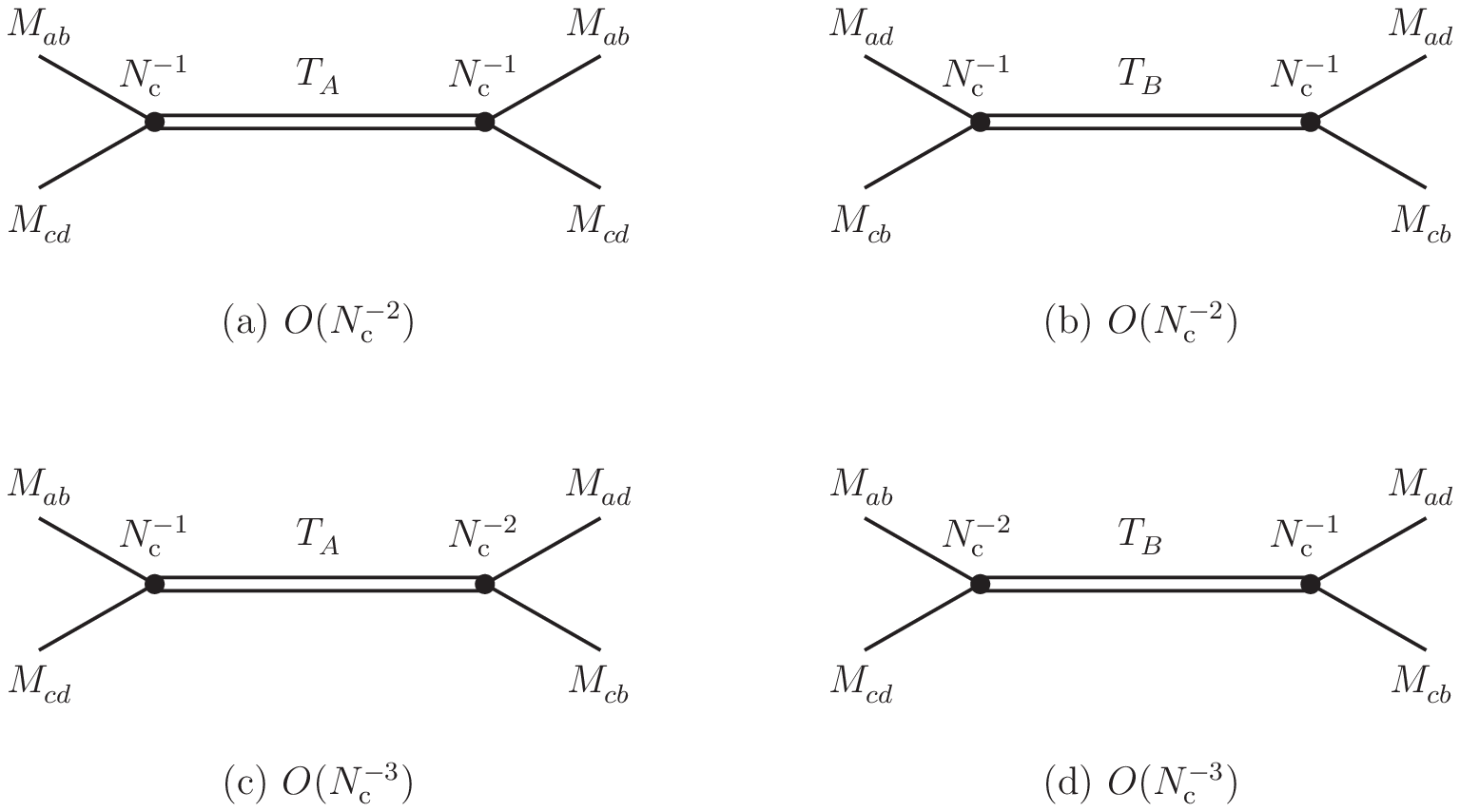,scale=0.75}
\caption{Leading-order contributions of tetraquarks $T_A^{}$ and 
$T_B^{}$ to the direct ((a) and (b)) and recombination ((c) and (d)) 
channels.}
\lb{2f7} 
\ec
\efg 
\par
Meson loops generally display ultraviolet divergences. These are
essentially absorbed in vertex and propagator renormalizations. 
The four-meson vertices being, in general, momentum dependent, new types 
of vertices may emerge, generating with higher-order loops an infinite 
series of terms in the meson effective Lagrangian. 
A typical example of that mechanism can be found in chiral perturbation
theory \cite{Gasser:1984gg}. For the purposes of the present approach
two remarks are of order. First, as can be observed from the previous
examples and diagrams, meson loops have weaker or at most equal 
dependences on $N_{\mathrm{c}}^{}$ than the generating four-meson
vertices. Therefore, vertex renormalizations cannot alter the leading
$N_{\mathrm{c}}^{}$-behaviors of existing tree-level vertices.
Second, since in our approach we are not displaying the detailed
momentum dependence of the vertices, the eventual appearance of new
types of vertices does not need the introduction of new tree-level
couplings, the existing ones representing generic types.  
\par    
The possible contributions of tetraquarks can be extracted from
Eqs. (\rf{2e6}) and (\rf{2e9}). It is evident that a single tetraquark 
alone cannot satisfy these two equations, unless one runs into 
contradictions. At least two different tetraquarks, which we denote
$T_A^{}$ and $T_B^{}$, are necessary to fulfill the above conditions.
The results are summarized as follows and in Fig. \rf{2f7}:
\bea
\lb{2e12}
& &A(T_A^{}\rightarrow M_{ab}^{}M_{cd}^{})\ \sim\ O(N_{\mathrm{c}}^{-1}),
\ \ \ \ \     
A(T_B^{}\rightarrow M_{ad}^{}M_{cb}^{})\ \sim\ O(N_{\mathrm{c}}^{-1}),\\
\lb{2e13}  
& &A(T_A^{}\rightarrow M_{ad}^{}M_{cb}^{})\ \sim\ O(N_{\mathrm{c}}^{-2}),
\ \ \ \ \     
A(T_B^{}\rightarrow M_{ab}^{}M_{cd}^{})\ \sim\ O(N_{\mathrm{c}}^{-2}).
\eea
\par
The decay widths of the tetraquarks are
\be \lb{2e14}
\Gamma(T_A) \sim\ \Gamma(T_B)\ =\ O(N_{\mathrm{c}}^{-2}),
\ee
which are smaller than those of the ordinary mesons 
($\Gamma=O(N_{\mathrm{c}}^{-1})$) by one power of $N_{\mathrm{c}}^{}$. 
\par 
The above properties provide us with an indication about the internal
structure of the tetraquark candidates. Transcribing the four-meson
couplings (\rf{2e10}) and (\rf{2e11}) [Figs. \rf{2f4}(a) and \rf{2f5}(a)]
into an effective interaction Lagrangian expressed by means of 
the corresponding quark color-singlet bilinears, one obtains
\bea \lb{2e15}
\mathcal{L}_{\mathrm{eff,int}}&=&-\frac{\lambda_1^{}}{N_c^{}}
[(\overline q_a^{}q_b^{})(\overline q_c^{}q_d^{})
(\overline q_d^{}q_a^{})(\overline q_b^{}q_c^{})+
(\overline q_a^{}q_d^{})(\overline q_c^{}q_b^{})
(\overline q_b^{}q_a^{})(\overline q_d^{}q_c^{})]\nonumber \\
& &-\frac{\lambda_2^{}}{N_c^{2}}
[(\overline q_a^{}q_b^{})(\overline q_c^{}q_d^{})
(\overline q_d^{}q_c^{})(\overline q_b^{}q_a^{})+
(\overline q_a^{}q_d^{})(\overline q_c^{}q_b^{})
(\overline q_b^{}q_c^{})(\overline q_d^{}q_a^{})],
\eea
where we have explicitly factored out the $N_\mathrm{c}^{}$-dependence 
of the coupling constants. One then deduces from Eqs. (\rf{2e12}) and 
(\rf{2e13}) that the tetraquark fields $T_A^{}$ and $T_B^{}$ should have 
the following structure in terms of the quark color-singlet bilinears:
\be \lb{2e16}
T_A^{}\ \sim\ (\overline q_a^{}q_d^{})(\overline q_c^{}q_b^{}),
\ \ \ \ \ \   
T_B^{}\ \sim\ (\overline q_a^{}q_b^{})(\overline q_c^{}q_d^{}), 
\ee
additional mixings between the two, of order $N_{\mathrm{c}}^{-1}$, 
being still possible.
\par
Manifestly, the above result favors a color singlet-singlet structure
of the tetraquarks in the exotic case. It is an open question 
whether the interquark confining forces may produce bound states
with such a structure.
\par
One might also encounter an intermediate situation, where one of the
tetraquarks, $T_B^{}$, say, is absent from the spectrum for some
dynamical reason. In that case, one tetraquark ($T_A^{}$) would exist
and, if the corresponding phase space is favorable, it would be observed 
through its preferred decay channel, as shown in (\rf{2e12}), 
\par

\section{Cryptoexotic channels} \lb{s3}
\setcounter{equation}{0}

We next consider cryptoexotic channels, with three different quark 
flavors, $a,b,c$, involved within the mesons $M_{ac}^{}$, $M_{cb}^{}$,
$M_{ab}^{}$ and $M_{cc}^{}$ [Eqs. (\rf{2e1}) and (\rf{2e2})]. Here
also, one may distinguish between direct (I and II) and 
recombination channels, described by the correlation functions
\par
\be 
\lb{3e1}
\Gamma_{\mathrm{I}}^{\mathrm{dir}}=\langle j_{ac}^{}j_{cb}^{}
j_{cb}^{\dagger}j_{ac}^{\dagger}\rangle,\ \ \ \   
\Gamma_{\mathrm{II}}^{\mathrm{dir}}=\langle j_{ab}^{}j_{cc}^{}
j_{cc}^{\dagger}j_{ab}^{\dagger}\rangle,\ \ \ \  
\Gamma^{\mathrm{rec}}=\langle j_{ac}^{}j_{cb}^{}j_{cc}^{\dagger}
j_{ab}^{\dagger}\rangle,\ \ \ \ \ \  a\neq b\neq c.
\ee
\par 
For the direct channel I, the leading and subleading diagrams
are represented in Fig. \rf{3f1}.
\par
\bfg 
\bc
\epsfig{file=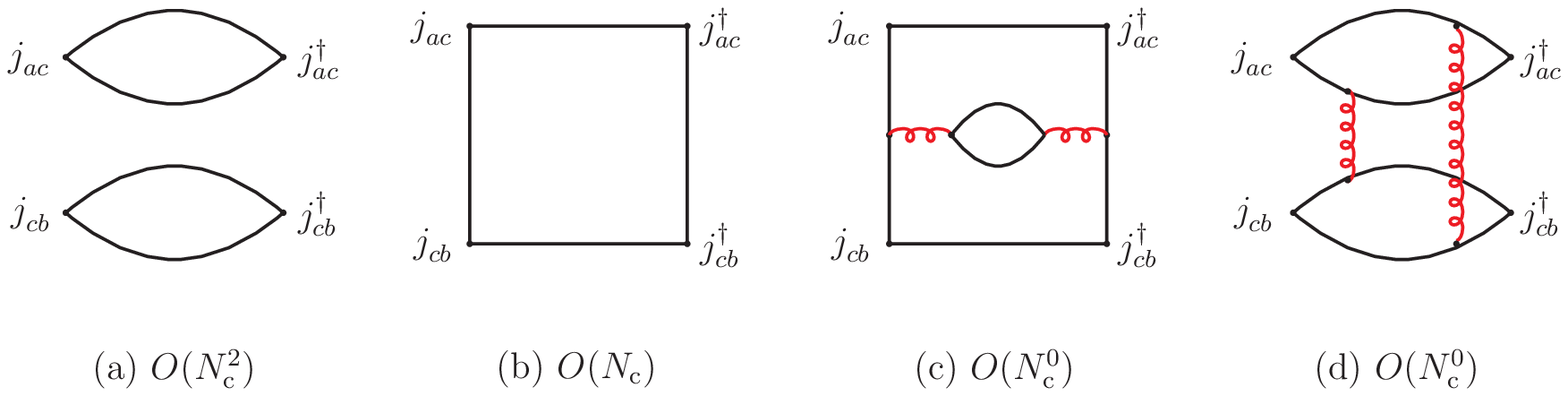,scale=0.75}
\caption{Leading- and subleading-order diagrams of the direct channel
I of (\rf{3e1}).}
\lb{3f1} 
\ec
\efg 
\par
Diagram (b) of Fig. \rf{3f1} represents the leading-order contribution
to the meson-meson scattering amplitude. It has only a two-quark 
singularity in the $s$-channel and therefore represents the
contribution of a single-meson intermediate state. Diagram (c)
represents contributions from radiative corrections to the previous
diagram. In its utmost left part, one has an intermediate state
made of a quark-antiquark pair and a single gluon, which later becomes 
a four-quark intermediate state and then retrieves back the former
situation. In the space of meson states, the first intermediate state
contributes to the formation of a single-meson state, which then emits 
two virtual mesons, or a tetraquark, and reabsorbs them later. Summing 
the chain of contributions of such
types of diagram, one ends up with a radiative correction to the 
one-meson propagator with a subleading order in $N_{\mathrm{c}}$. 
This diagram may also describe a mixing between a single-meson state
and a tetraquark state, having the same quantum numbers. We shall
come back to this question when discussing the mixing problem in the 
recombination channel.
\par
Diagram (d) represents a direct contribution of two-meson states and/or 
of a tetraquark state. One obtains the following leading-order behaviors 
for the two-meson scattering amplitudes and the transition amplitudes
through two-meson and tetraquark intermediate states:
\bea
\lb{3e2}
& &A(M_{ac}^{}M_{cb}^{}\rightarrow M_{ac}^{}M_{cb}^{})\ 
=\ O(N_{\mathrm{c}}^{-1}),\\
\lb{3e3}
& &A(M_{ac}^{}M_{cb}^{}\rightarrow MM\rightarrow 
M_{ac}^{}M_{cb}^{})\ =\ O(N_{\mathrm{c}}^{-2}),\\
\lb{3e4}
& &A(M_{ac}^{}M_{cb}^{}\rightarrow T\rightarrow M_{ac}^{}M_{cb}^{})\ 
=\ O(N_{\mathrm{c}}^{-2}).
\eea
\par
For the direct channel II, the structure of the diagrams is similar to
that of Fig. \rf{2f1}, represented in Fig. \rf{3f2}, from which one deduces
\bea
\lb{3e5}
& &A(M_{ab}^{}M_{cc}^{}\rightarrow M_{ab}^{}M_{cc}^{})\ 
=\ O(N_{\mathrm{c}}^{-2}),\\
\lb{3e6}
& &A(M_{ab}^{}M_{cc}^{}\rightarrow MM\rightarrow 
M_{ab}^{}M_{cc}^{})\ =\ O(N_{\mathrm{c}}^{-2}),\\
\lb{3e7}
& &A(M_{ab}^{}M_{cc}^{}\rightarrow T\rightarrow M_{ab}^{}M_{cc}^{})\ 
=\ O(N_{\mathrm{c}}^{-2}).
\eea
\par
\bfg 
\bc
\epsfig{file=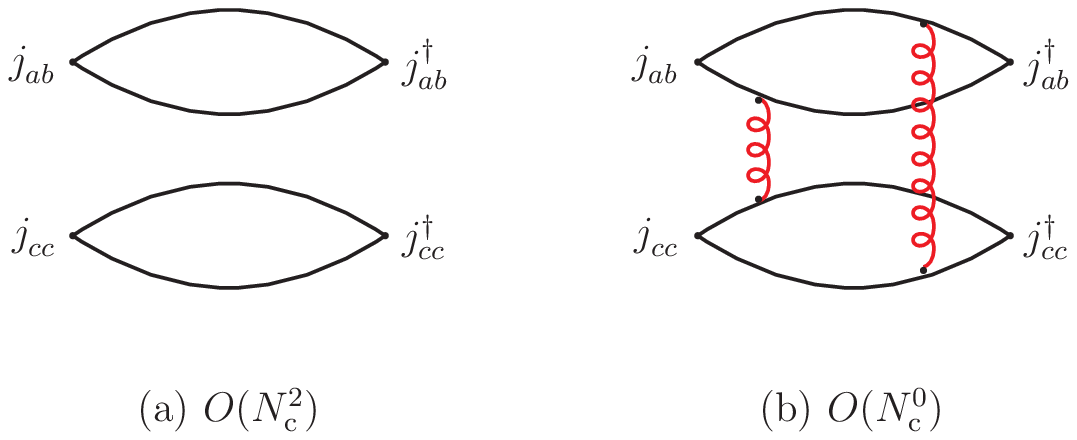,scale=0.75}
\caption{Leading- and subleading-order diagrams of the direct channel
II of (\rf{3e1}).}
\lb{3f2} 
\ec
\efg 
\par
For the recombination channel of (\rf{3e1}), the main leading and 
subleading diagrams are shown in Fig. \rf{3f3}. (The diagram
similar to that of Fig. \rf{2f2}(d) is not drawn, since it 
contributes to subleading radiative corrections in the 
$u$-channel.) Diagram (a) does not have $s$-channel singularities
(cf. Figs. \rf{2f2}(a) and \rf{2f5}(a) and the appendix), while diagrams
(b) and (c) receive contributions from four-quark intermediate
states in the $s$-channel (cf. the appendix).
\bfg 
\bc
\epsfig{file=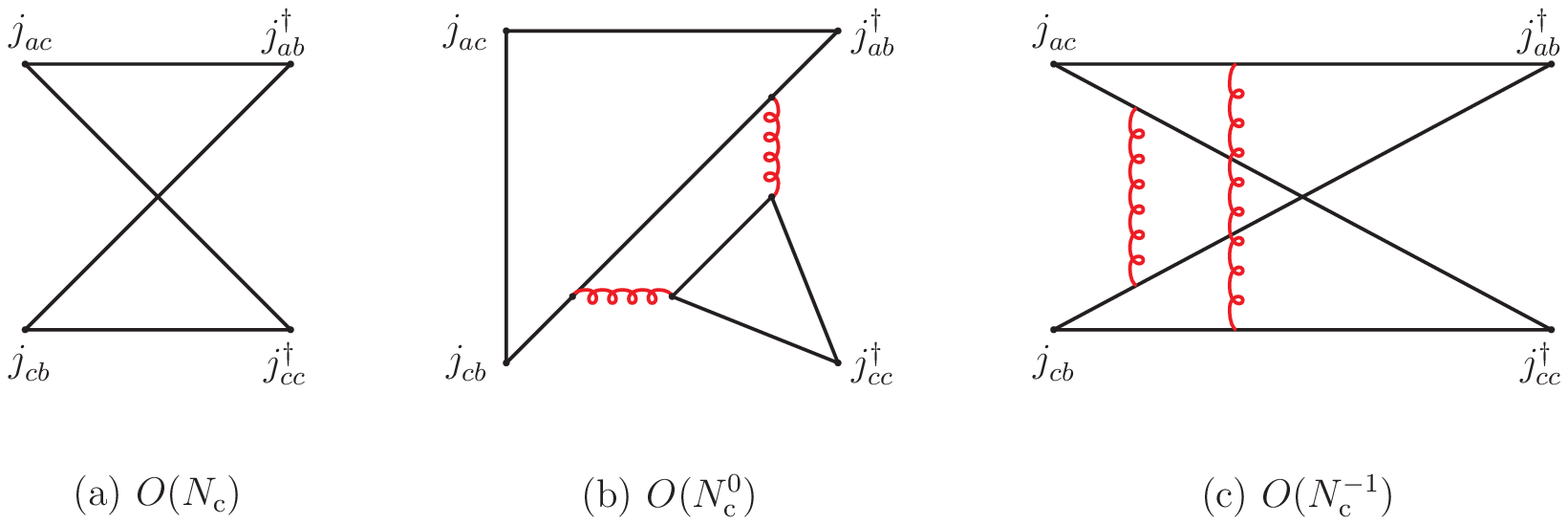,scale=0.75}
\caption{Leading- and subleading-order diagrams of the recombination 
channel of (\rf{3e1}).}
\lb{3f3} 
\ec
\efg 
\par
One then deduces the following properties of the meson-meson
scattering amplitudes and the transition amplitudes
through two-meson and tetraquark intermediate states:
\bea
\lb{3e8}
& &A(M_{ac}^{}M_{cb}^{}\rightarrow M_{ab}^{}M_{cc}^{})\ 
=\ O(N_{\mathrm{c}}^{-1}),\\
\lb{3e9}
& &A(M_{ac}^{}M_{cb}^{}\rightarrow MM\rightarrow 
M_{ab}^{}M_{cc}^{})\ =\ O(N_{\mathrm{c}}^{-2}),\\
\lb{3e10}
& &A(M_{ac}^{}M_{cb}^{}\rightarrow T\rightarrow M_{ab}^{}M_{cc}^{})\ 
=\ O(N_{\mathrm{c}}^{-2}).
\eea
\par
The information obtained about the $N_{\mathrm{c}}^{}$-behaviors of 
leading and subleading diagrams can now be transcribed into properties 
of effective meson-meson interactions. These are summarized in Fig.
\rf{3f4}.
\bfg 
\bc
\epsfig{file=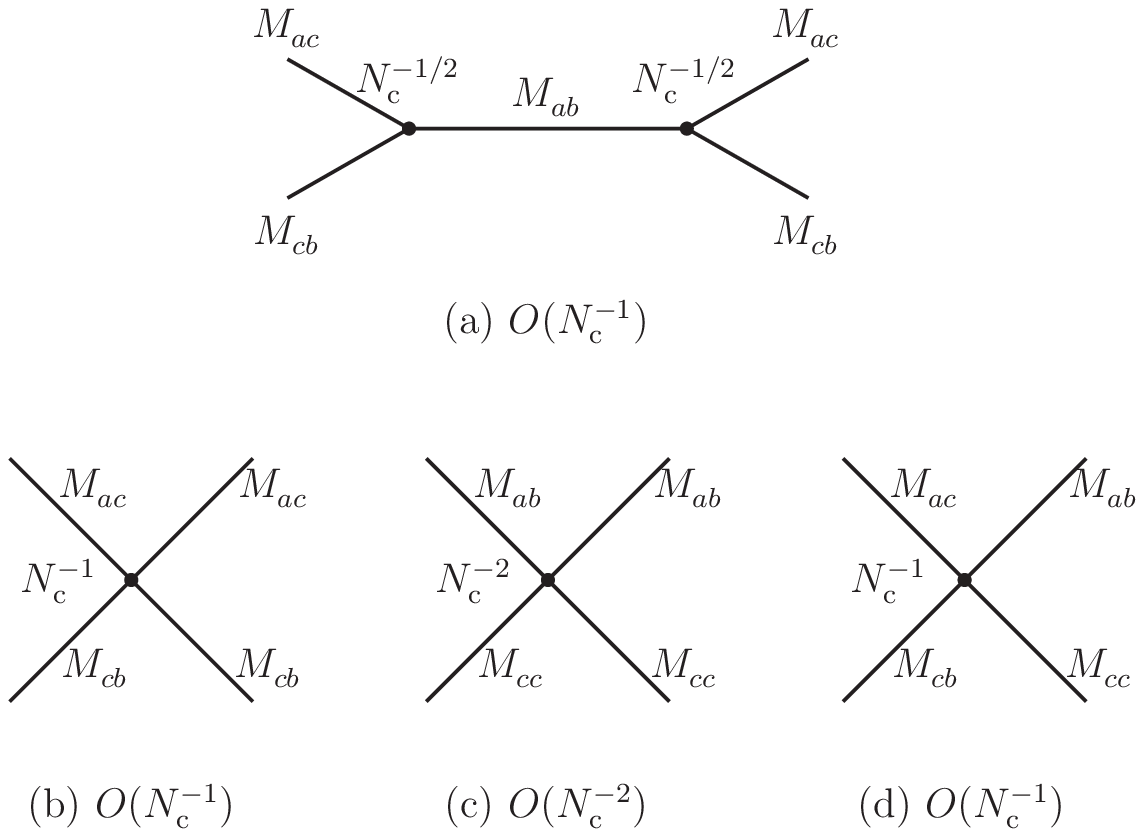,scale=0.75}
\caption{Significant tree and vertex diagrams with meson propagators in 
the direct channel I [(a) and (b)], the direct channel II [(c)] and
the recombination channel [(d)].} 
\lb{3f4} 
\ec
\efg 
\par  
One can then evaluate the contributions of two-meson intermediate 
states in the scattering amplitudes. The results, for the leading
terms, are presented in Fig. \rf{3f5}. They manifestly reproduce
the $N_{\mathrm{c}}^{}$-behaviors as expected from Eqs. (\rf{3e3}),
(\rf{3e6}) and (\rf{3e9}). 
\par
\bfg 
\bc
\epsfig{file=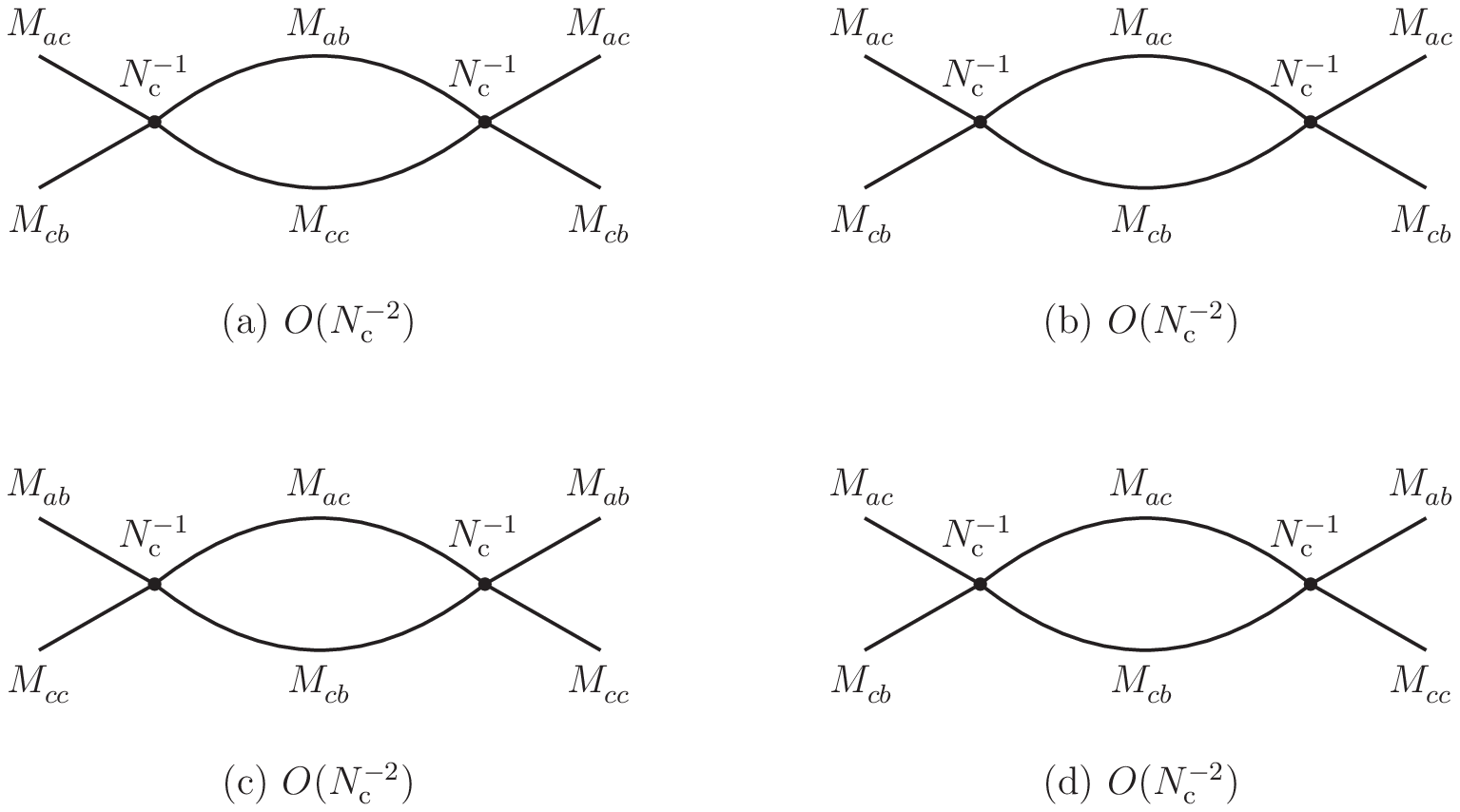,scale=0.75}
\caption{Two-meson intermediate-state contributions to the three channels
at $N_{\mathrm{c}}^{}$-leading order: direct channel I [(a) and (b)], 
direct channel II [(c)] and recombination channel [(d)].} 
\lb{3f5} 
\ec
\efg 
\par  
The tetraquark contributions can be extracted in the same way as for 
the exotic channels. Because of the presence of the additional diagram
(b) of Fig. \rf{3f3}, they are of the same order in all three channels
and hence a single tetraquark $T$ can accommodate all the
corresponding constraints. The results are summarized in Fig. \rf{3f6}.    
The decay width of the tetraquark is again of order 
$N_{\mathrm{c}}^{-2}$:
\be \lb{3e11}
\Gamma(T)=O(N_{\mathrm{c}}^{-2}).
\ee
\bfg 
\bc
\epsfig{file=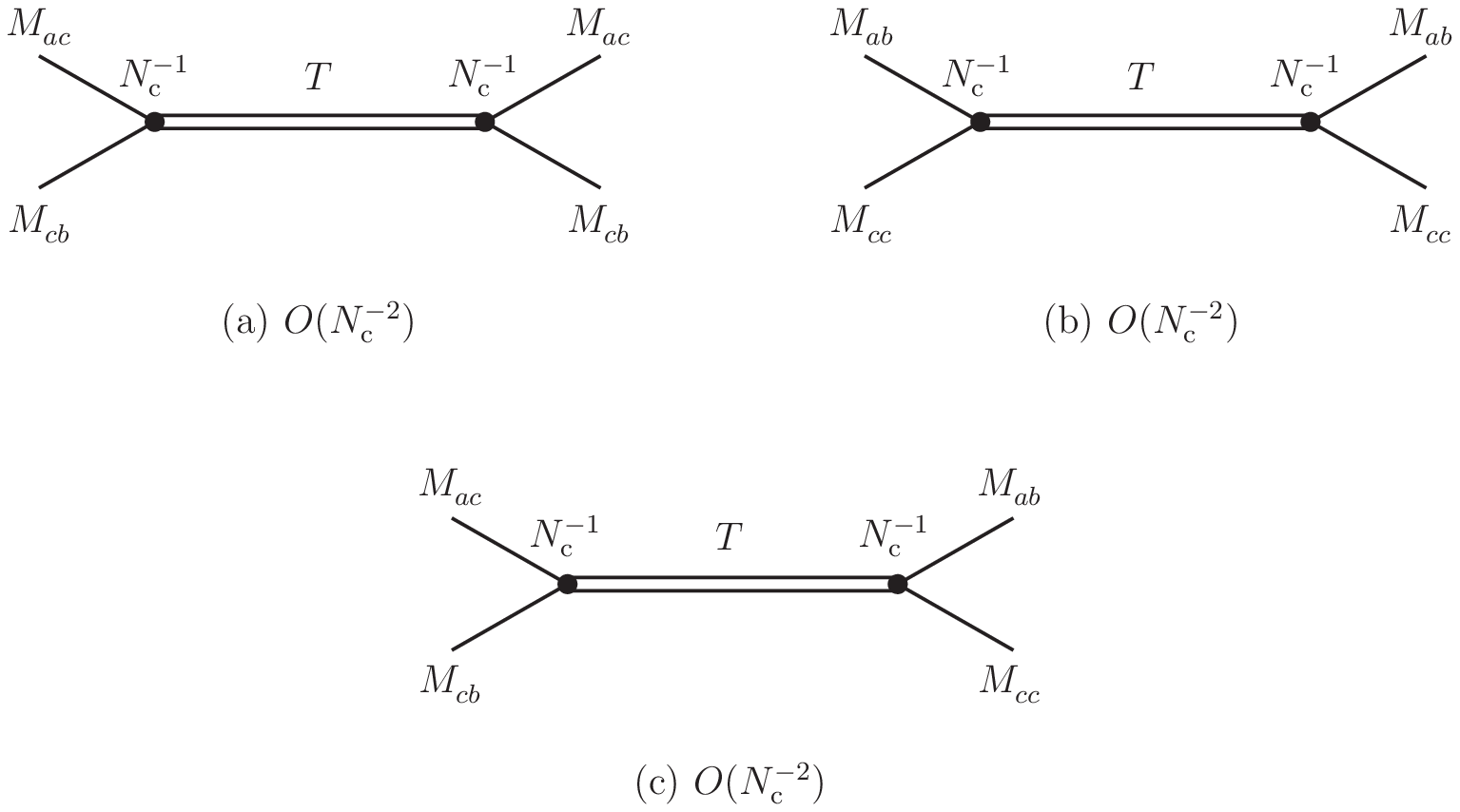,scale=0.75}
\caption{Tetraquark-state contributions to the three channels
at $N_{\mathrm{c}}^{}$-leading order: direct channel I [(a)], direct 
channel II [(b)] and recombination channel [(c)].} 
\lb{3f6} 
\ec
\efg 
\par  
Actually, diagram (b) of Fig. \rf{3f3} may also describe a mixing of 
two-meson states or of a tetraquark state with a single-meson state
that appears in the left part of the diagram. The corresponding
mixings, which do not change the previous results, are described in 
Fig. \rf{3f7}.
\bfg 
\bc
\epsfig{file=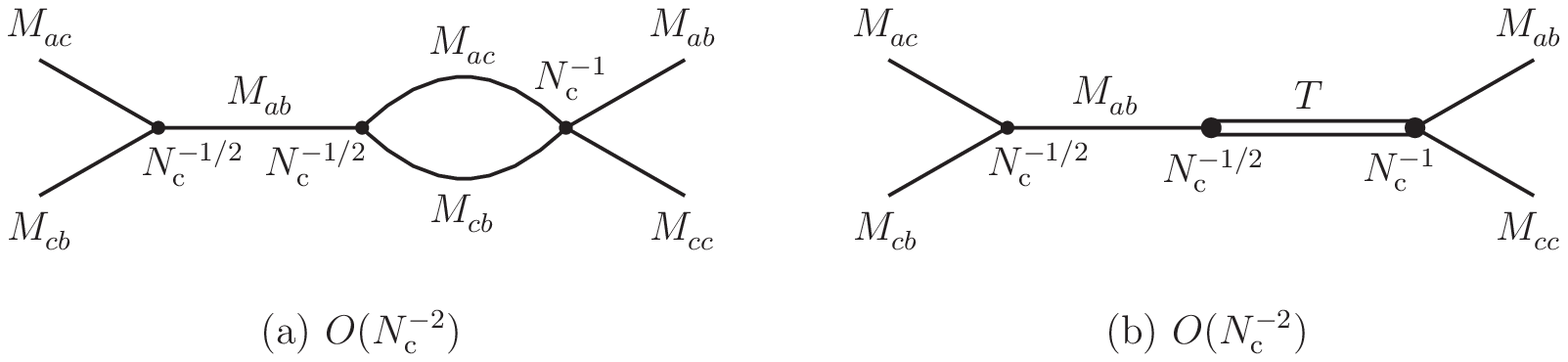,scale=0.75}
\caption{Mixings, in the recombination channel, of a single meson state
with two-meson [(a)] and tetraquark [(b)] states.} 
\lb{3f7} 
\ec
\efg 
\par
Meson-tetraquark mixings also exist in the direct channel I, as was 
previously mentioned, emerging from diagrams of the type of Fig.
\rf{3f1}(c). Here, since the quark and its antiquark can be created 
with any flavor, the resulting two-meson and/or tetraquark states may 
belong to another class of cryptoexotic states. The corresponding 
mixings are described in Fig. \rf{3f8}. They do not change the leading 
coupling properties of cryptoexotic tetraquarks to two-meson states. 
\bfg 
\bc
\epsfig{file=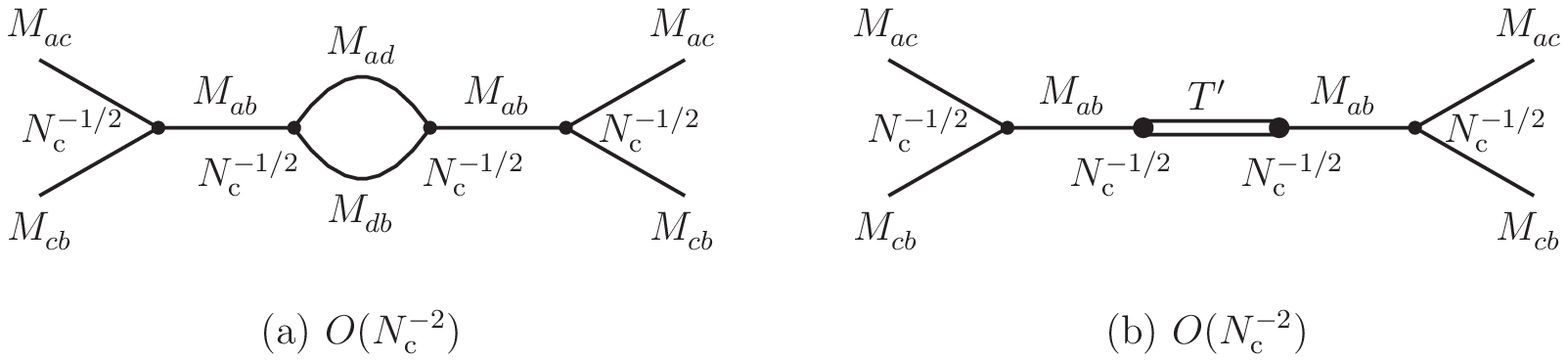,scale=0.75}
\caption{Mixings, in the direct channel I, of a single meson state
with two-meson [(a)] and tetraquark [(b)] states. The flavor of the
internal loop quark is designated by $d$ and the corresponding 
tetraquark by $T'$.} 
\lb{3f8} 
\ec
\efg 
\par
Cryptoexotic tetraquarks have therefore the possibility of decaying 
into two mesons either through direct coupling or through mixing with
single-meson states. Both types of decay lead to the same 
$O(N_{\mathrm{c}}^{-1})$ behavior of the corresponding transition
amplitude.   
\par
The fact that a single tetraquark satisfies all the existing constraints,
does not exclude the possibility of the existence of two types of 
tetraquark, with different internal structures. However, contrary to the 
exotic case [Eqs. (\rf{2e12}) and (\rf{2e13})], these would not have 
now preferred decay channels. A particular case may emerge when, for 
some dynamical reason, tetraquarks do not contribute to diagram (b) of 
Fig. \rf{3f3}. In that case, one falls back into the situation of the 
exotic case, where two different tetraquarks, each one having a preferred 
decay channel, are needed.
\par 
%\newpage
We now consider the case of the open-type channel, where the quark 
flavor $c$ appears through two quark fields, rather than a pair of a 
quark and an antiquark field. The four-point correlation function 
describing the corresponding meson-meson scattering is
\be 
\lb{3e12}
\Gamma=\langle j_{ac}^{}j_{bc}^{}j_{bc}^{\dagger}j_{ac}^{\dagger}
\rangle.
\ee
Here, the direct and recombination channels are identical, with the
common process $M_{ac}^{}M_{bc}^{}\rightarrow M_{ac}^{}M_{bc}^{}$.   
The corresponding $N_{\mathrm{c}}^{}$-leading and subleading diagrams are 
represented in Fig. \rf{3f9}.  
\bfg 
\bc
\epsfig{file=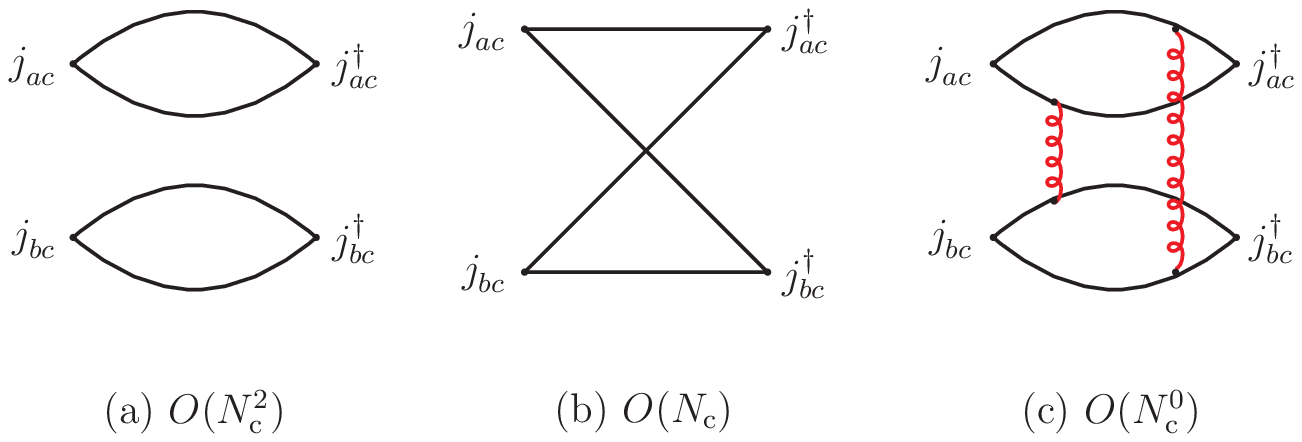,scale=0.75}
\caption{Leading and subleading diagrams of the correlation function
(\rf{3e12}).}
\lb{3f9} 
\ec
\efg 
\par
Their implication in terms of the four-meson vertex and two-meson
and tetraquark intermediate states is shown in Fig. \rf{3f10}.
\bfg 
\bc
\epsfig{file=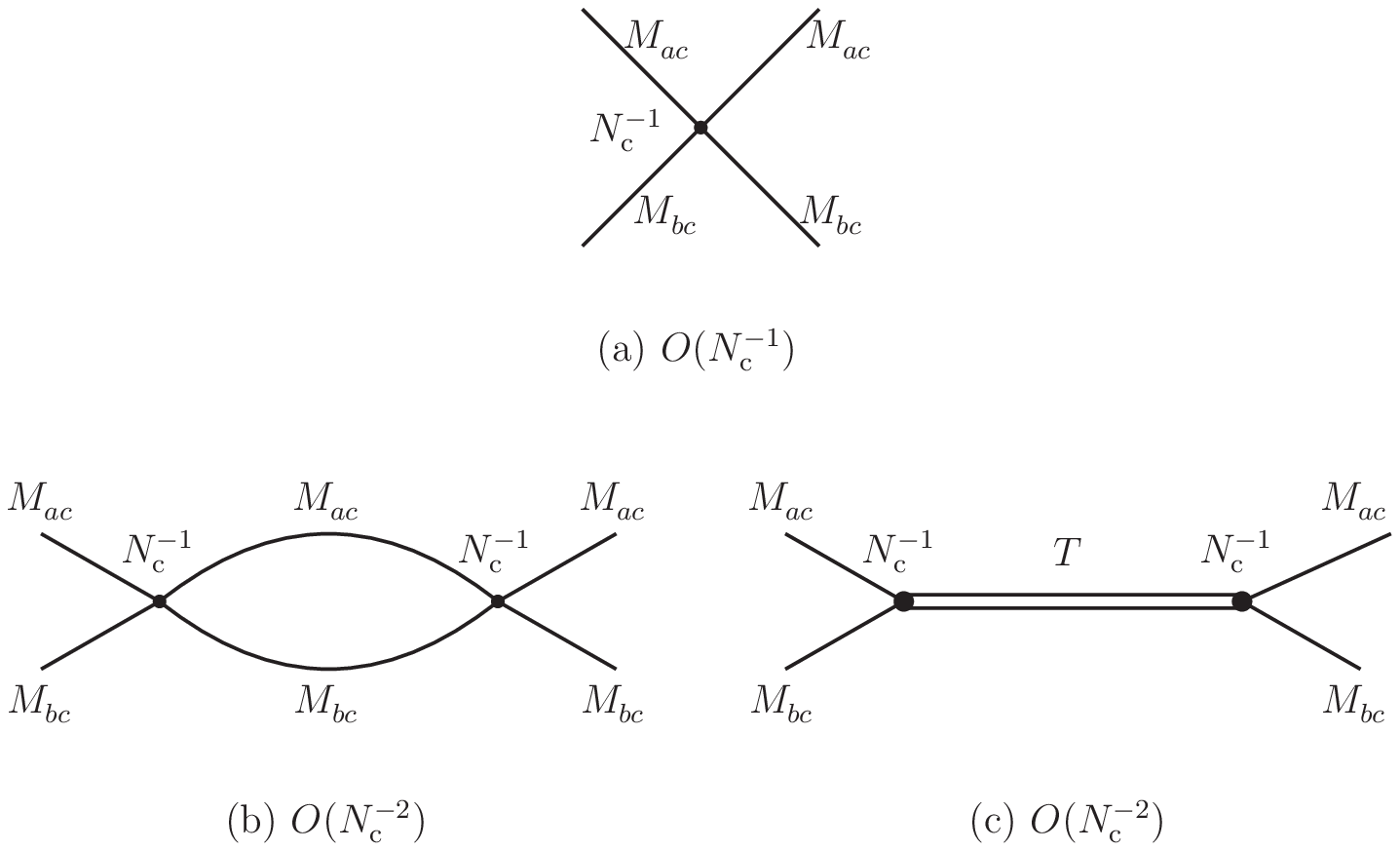,scale=0.75}
\caption{Four-meson vertex [(a)], two-meson intermediate states [(b)]
and tetraquark state [(c)] arising from the correlation function
(\rf{3e12}).}
\lb{3f10} 
\ec
\efg 
\par
One also obtains a decay width of the order of $N_{\mathrm{c}}^{-2}$ 
[Eq. (\rf{3e11})]. 
\par
The case of cryptoexotic channels with two quark flavors can be
treated in the same way as for the case with three flavors. 
The corresponding correlation functions are
\be 
\lb{3e13}
\Gamma_I^{\mathrm{dir}}=\langle j_{ac}^{}j_{ca}^{}j_{ca}^{\dagger}
j_{ac}^{\dagger}\rangle,\ \ \ \ \ \ \  
\Gamma_{II}^{\mathrm{dir}}=\langle j_{aa}^{}j_{cc}^{}j_{cc}^{\dagger}
j_{aa}^{\dagger}\rangle, \ \ \ \ \ \ \
\Gamma^{\mathrm{rec}}=\langle j_{ac}^{}j_{ca}^{}j_{cc}^{\dagger}
j_{aa}^{\dagger}\rangle,\ \ \ \ \ \ \  a\neq c.
\ee
\par 
Most of the relevant diagrams are similar to those found in the 
three-flavor case. In addition, one finds annihilation-type diagrams 
involving gluon lines. In particular, the direct channel I of (\rf{3e13})
involves, among others, an annihilation diagram with two gluon lines
(similar to diagram  (c) of Fig. \rf{3f9}, rotated by $\pi/2$), which
produces, in meson space, a glueball as an intermediate state in the 
$s$-channel (the analog of Fig. \rf{2f4}(b) in the $s$-channel).
The presence of the new diagrams does not change the qualitative results 
found earlier in the large-$N_{\mathrm{c}}^{}$ limit. Mixings of 
tetraquarks with glueball states are of subleading order. Therefore, 
the main conclusions about the tetraquark decay width and two-meson 
intermediate states remain unchanged.
\par

\section{Conclusion} \lb{s4}
\setcounter{equation}{0}

The study of compact tetraquark properties in the 
large-$N_{\mathrm{c}}^{}$ limit of QCD through the meson-meson 
scattering amplitudes involves $N_{\mathrm{c}}^{}$-subleading diagrams, 
at which order also two-meson states occur. It was shown that the latter 
consistently satisfy the constraints emerging from the 
$1/N_{\mathrm{c}}^{}$ expansion procedure, thus providing a firm basis 
for the extraction of the tetraquark properties. Considering many types 
of quark flavor and various combinations of mesons in the $s$-channel, 
it was found that, in general, tetraquarks have narrow decay widths, of 
the order of $N_{\mathrm{c}}^{-2}$, much smaller than those of ordinary 
mesons. For the particular case of exotic tetraquarks, involving four 
different quark flavors, two different types of tetraquark are needed 
to satisfy the consistency constraints, each of them having a 
preferred decay channel and a quark structure made of the
product of two color-singlet bilinears.
\par
In the real world, where $N_{\mathrm{c}}^{}=3$, some of the qualitative 
aspects found above might be attenuated. Nevertheless, they might
still serve as a guidance for quantitative investigations. 
\par 
\vspace{1 cm}
%\newpage

\noindent
{\Large \textbf{Acknowledgements}}
\par
D.~M.~acknowledges support from the Austrian Science Fund (FWF),
Grant No. P29028.
The figures were drawn with the aid of the package Axodraw
\cite{Vermaseren:1994je}.
\par
%\newpage

\appendix
\renewcommand{\theequation}{\Alph{section}.\arabic{equation}}
\renewcommand{\thesection}{\Alph{section}.}

\section{Landau equations} \lb{a1}
\setcounter{equation}{0}

For completeness, we present in this appendix several typical
cases, where the Landau equations are used for the determination 
of the location of singularities occurring in Feynman diagrams involved
in the large-$N_{\mathrm{c}}^{}$ limit. A generic expression of the latter is
\be \lb{ea1}
I(p)=\int\prod_{\ell=1}^{L}\frac{d^4q_{\ell}^{}}{(2\pi)^4}
\prod_{i=1}^{I}\frac{1}{(k_i^2-m_i^2+i\epsilon)},
\ee
where $p$ represents a collection of external momenta and $k_i$ are linear 
functions of the $p$'s and of the loop variables $q$.
\par
The Landau equations are \cite{Landau:1959fi,Itzykson:1980rh}
\bea
\lb{ea2}
& &\lambda_i^{}(k_i^2-m_i^2)=0,\ \ \ \ \ \ \ i=1,\ldots,I,\\
\lb{ea3}
& &\sum_{i=1}^I\lambda_i k_i^{}\cdot\frac{\partial k_i^{}}
{\partial q_{\ell}^{}}=0,\ \ \ \ \ \ \ \ \ell=1,\ldots,L,
\eea  
where the $\lambda$ are parameters (Lagrange multipliers) to be
determined.
\par
This system of equations may have independent subsystems, corresponding
to the vanishing of a certain number of parameters $\lambda$.
\par
We are mainly interested in the singularities produced by the quark
propagators. Gluon propagators may also produce singularities, but
since the gluons are massless, and except in annihilation diagrams,
these generally start at the same positions as those produced by
the quark propagators. We therefore will not consider gluon
propagators in the Landau equations, except in one case, for 
illustrative purposes; this amounts to taking the corresponding 
$\lambda$ equal to zero.
\par
However, gluon lines may play a decisive role in the production of
quark singularities through the momentum they carry. We can
distinguish two types of gluon contribution. For the first category,
the gluon lines participate in the renormalization of existing objects,
like propagators or vertices. In this case, one could ignore them
for the present purpose.
For the second category, they participate in the interaction 
between different clusters of objects and play an active role in
the formation of singularities related to intermediate states. Those
should be considered, when present.
\par
We shall mainly consider recombination-type diagrams of the exotic 
channels.
\par
We begin with the recombination diagram of leading order 
(Figs. \rf{2f2}(a) and \rf{af1}).
\bfg 
\vspace*{1 cm}
\bc
\epsfig{file=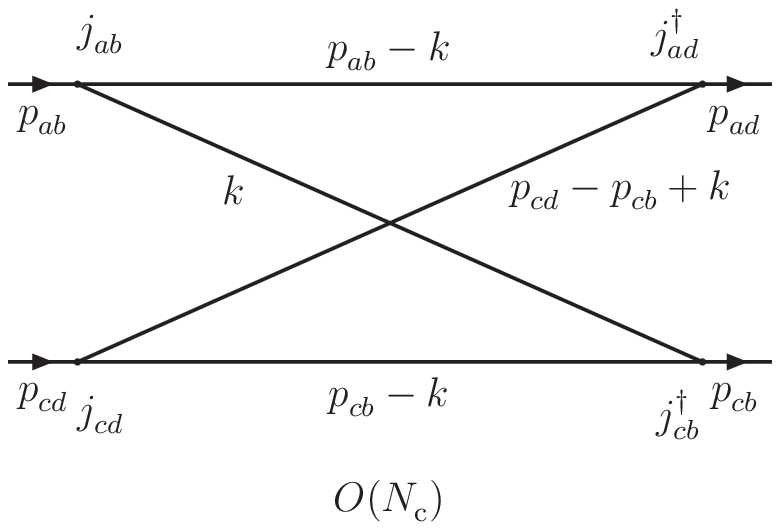,scale=0.75}
\caption{Recombination diagram of Fig. \rf{2f2}(a) with explicit
momentum flow.}
\lb{af1} 
\ec
\efg
The external momenta are $p_{ab}^{}$, $p_{cd}^{}$, $p_{ad}^{}$,
$p_{cb}^{}$, with the definitions
\be \lb{ea4}
P=p_{ab}^{}+p_{cd}^{}=p_{ad}^{}+p_{cb}^{},\ \ \ \ \ 
s=P^2,\ \ \ \ \ t=(p_{ab}^{}-p_{ad}^{})^2,\ \ \ \ \ \
u=(p_{ab}^{}-p_{cb}^{})^2.
\ee 
\par
The Landau equations are
\bea
\lb{ea5}
& &\lambda_a^{}((p_{ab}^{}-k)^2-m_a^2)=0,\ \ \ \ \ \      
\lambda_b^{}(k^2-m_b^2)=0,\nonumber \\
& &\lambda_c^{}((p_{cb}^{}-k)^2-m_c^2)=0,\ \ \ \ \ \  
\lambda_d^{}((p_{cd}^{}-p_{cb}^{}+k)^2-m_d^2)=0,\\
\lb{ea6}
& &-\lambda_a^{}(p_{ab}^{}-k)+\lambda_b^{}k-\lambda_c^{}
(p_{cb}^{}-k)+\lambda_d^{}(p_{cd}^{}-p_{cb}^{}+k)=0.
\eea
\par
This system of equations has several independent subsystems.
Choosing $\lambda_b^{}=\lambda_d^{}=0$, one obtains 
$u=(m_a^{}\pm m_c^{})^2$.
(Only physical singularities, corresponding to $+$ signs between the 
masses, are relevant.)
Choosing $\lambda_a^{}=\lambda_c^{}=0$, one obtains 
$t=(m_b^{}\pm m_d^{})^2$.
Choosing $\lambda_c^{}=\lambda_d^{}=0$, one obtains 
$p_{ab}^2=(m_a\pm m_b)^2$, and so forth. The latter type of singularities 
are present inside the meson propagators. 
The property that the singularities in $u$ and $t$
involve only two quark masses is reminiscent of the fact that there 
are no four-quark singularities. The $u$- and $t$-channel singularities
will be saturated by one-meson intermediate states. No singularities in 
$s$ are found.
\par
The general system involving the four $\lambda$ leads to an
equation where $u$, $t$ and the $p^2$ enter; it comes from the
condition of the vanishing of the determinant of a $4\times 4$
matrix. In principle, this equation should give information about
the possible existence of anomalous thresholds or unphysical 
thresholds in the $u$- and $t$-channels, depending on the values 
taken by the $p^2$. The cases of interest are those where the 
$p^2$ represent masses squared of the external mesons.
The equation considerably simplifies in the equal-mass case. Taking 
$m_a=m_b=m_c=m_d=m$ and $p_{ab}^2=p_{cd}^2=p_{ad}^2=p_{cb}^2=p^2$, one 
finds that the singularities occur on the unphysical sheets, at $u=0$ or 
$t=0$, which are the equal-mass limits of $u=(m_a-m_c)^2$ and 
$t=(m_b-m_d)^2$, respectively. This shows the general tendancy of the 
equation of not producing anomalous thresholds on the physical sheet. 
In any event, no $s$-channel singularities arise. 
\par
The next example contains one-gluon exchange between quarks $a$ and $b$
(Fig. \rf{af2}).
\bfg 
%\vspace*{1 cm}
\bc
\epsfig{file=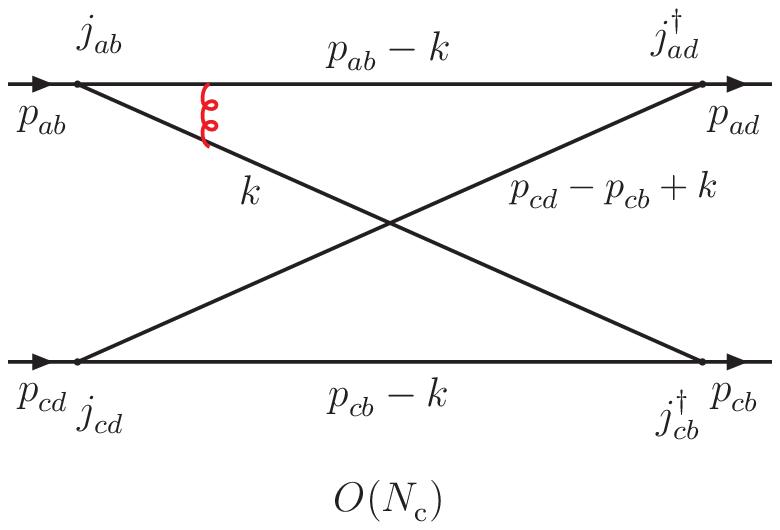,scale=0.75}
\caption{One-gluon exchange between quarks $a$ and $b$.}
\lb{af2} 
\ec
\efg
\par
Here, the loop momenta can be chosen in such a way that the quark
propagators, cut by a vertical line on the right of the gluon 
propagator, have the same momentum dependence as in the preceding
example. This shows, as expected, that the gluon does not play any
role for the determination of the $s$- or $t$- or $u$-channel 
singularities. It contributes to the renormalization of the current
vertex or to the incoming meson propagator. Cutting the diagram with
a vertical line on the left of the gluon propagator does not lead to 
new singularities.  
\par
The third example contains one-gluon exchange between quarks $a$ and
$d$ (Fig. \rf{af3}).
\bfg 
%\vspace*{1 cm}
\bc
\epsfig{file=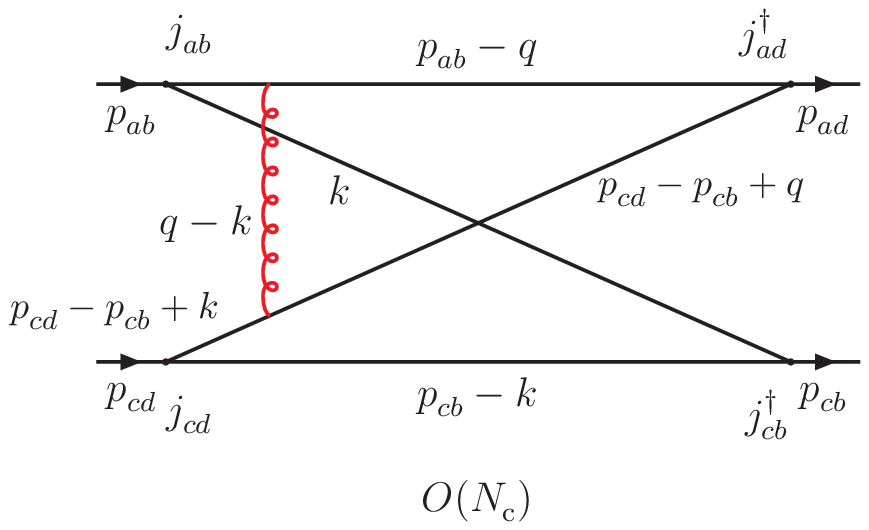,scale=0.75}
\caption{One-gluon exchange between quarks $a$ and $d$.}
\lb{af3} 
\ec
\efg
\par
Taking the vertical cut on the right of the gluon line, one obtains
the equations
\bea
\lb{ea7}
& &\lambda_a^{}((p_{ab}^{}-q)^2-m_a^2)=0,\ \ \ \ \ \      
\lambda_b^{}(k^2-m_b^2)=0,\nonumber \\
& &\lambda_c^{}((p_{cb}^{}-k)^2-m_c^2)=0,\ \ \ \ \ \  
\lambda_d^{}((p_{cd}^{}-p_{cb}^{}+q)^2-m_d^2)=0,\\
\lb{ea8}
& &-\lambda_a^{}(p_{ab}^{}-q)+\lambda_d^{}(p_{cd}^{}-p_{cb}^{}+q)=0,
\ \ \ \ \ \ \lambda_b^{}k-\lambda_c^{}(p_{cb}^{}-k)=0.
\eea
\par
The system of equations factorizes into two independent subsystems,
with solutions $p_{ad}^2=(m_a^{}\pm m_d^{})^2$ and
$p_{cb}^2=(m_b^{}\pm m_c^{})^2$.
\par
Taking the vertical cut on the left of the gluon line, one finds the 
same equations as for Fig. \rf{af1}, with the same solutions.
\par
For completeness, we also consider the case of a gluon exchanged
between quarks $b$ and $d$. The case where the gluon line is not cut 
by the vertical line, corresponds to a figure similar to Fig. \rf{af3}. 
With a relabeling of momenta, one obtains the same type of Landau 
equations, with similar conclusions. For illustration, we consider 
here the case where the gluon line is cut by the vertical line
[Fig. \rf{af4}]. 
\bfg
%\vspace*{1 cm}
\bc
\epsfig{file=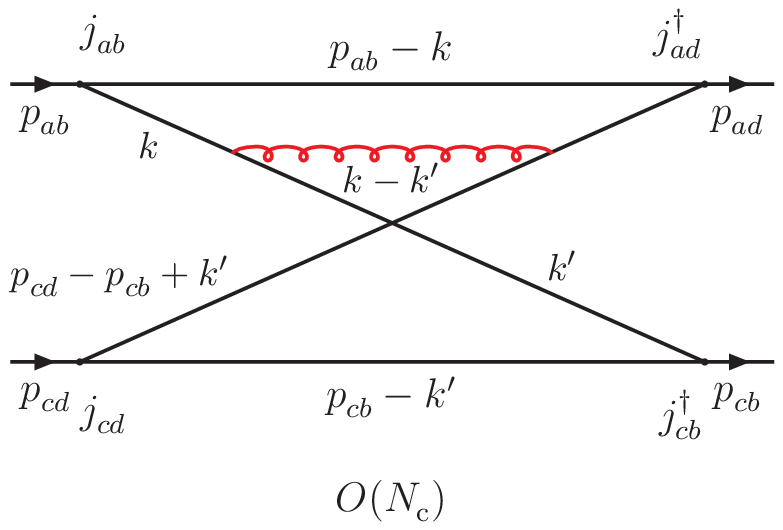,scale=0.75}
\caption{One-gluon exchange between quarks $b$ and $d$.}
\lb{af4} 
\ec
\efg
\par
The Landau equations are:
\bea
\lb{ea9}
& &\lambda_a^{}((p_{ab}^{}-k)^2-m_a^2)=0,\ \ \ \ \  
\lambda_b^{}(k^{\prime 2}-m_b^2)=0,\nonumber \\
& &\lambda_c^{}((p_{cb}^{}-k')^2-m_c^2)=0,\ \ \ \ \
\lambda_g((k-k')^2-m_g^2)=0,\nonumber \\
& &\lambda_d^{}((p_{cd}^{}-p_{cb}^{}+k')^2-m_d^2)=0, \\
\lb{ea10}
& &-\lambda_a^{}(p_{ab}^{}-k)+\lambda_g^{}(k-k')=0,
\nonumber \\
& &-\lambda_g(k-k')+\lambda_b^{}k'-\lambda_c^{}(p_{cb}^{}-k')
+\lambda_d(p_{cd}-p_{cb}+k')=0,
\eea
where we have also attributed a mass $m_g$ to the gluon, in order
to have a more explicit control of its effect.
The nontrivial solution of these equations corresponds to the
case $\lambda_b=\lambda_d=0$, leading to a singularity at
$u=(m_a+m_c+m_g)^2$. One verifies that the gluon mass contributes
additively to the quark masses in the existing $u$-channel singularity.
In the limit where the gluon mass tends to zero, the corresponding
threshold tends to the existing two-quark threshold.   
\par
The fifth example contains two-gluon exchanges between quarks 
$a$ and $c$, and $b$ and $d$, respectively [Fig. \rf{af5}]. 
\bfg 
%\vspace*{1 cm}
\bc
\epsfig{file=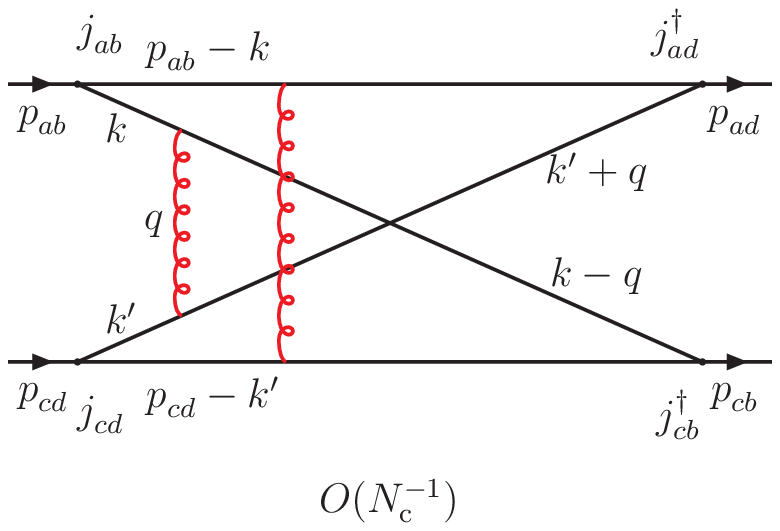,scale=0.75}
\caption{Two-gluon exchanges between the four quarks.}
\lb{af5} 
\ec
\efg
\par
The vertical cut is taken between the two gluon lines. The Landau
equations are
\bea
\lb{ea11}
& &\lambda_a^{}((p_{ab}^{}-k)^2-m_a^2)=0,\ \ \ \ \ \      
\lambda_b^{}((k-q)^2-m_b^2)=0,\nonumber \\
& &\lambda_c^{}((p_{cd}^{}-k')^2-m_c^2)=0,\ \ \ \ \ \  
\lambda_d^{}((k'+q)^2-m_d^2)=0,\\
\lb{ea12}
& &-\lambda_a^{}(p_{ab}^{}-k)+\lambda_b^{}(k-q)=0,\ \ \ \ \ 
-\lambda_c^{}(p_{cd}^{}-k')+\lambda_d^{}(k'+q)=0,\nonumber \\ 
& &-\lambda_b^{}(k-q)+\lambda_d^{}(k'+q)=0.
\eea
\par
The system of equations can be solved: one finds the physical
singularity at $P^2=s=(m_a^{}+m_b^{}+m_c^{}+m_d^{})^2$. (The unphysical
singularities, corresponding to changes of sign in front of the 
masses, exist as well.) The fact that four masses are present means
that we have four-quark intermediate states, which then would
generate, together with the contributions of other diagrams involving 
increasing numbers of gluon exchanges, two-meson and eventually tetraquark
states.
\par
The reason of the appearance of the $s$-channel singularity is 
related to the iterative nature of the diagram. Considering, in general,
the four-quark Green's function (eight-point function), the diagrams
describing the latter are usually divided into irreducible
and reducible types, the latter being obtained by iteration of the 
former. In the present case, the diagram is one of the first iterations 
of one-gluon exchange diagrams.
\par
We next consider the recombination diagram (b) of Fig. \rf{3f3} of
the cryptoexotic channels (Fig. \rf{af6}), involving three different
quark flavors.
\bfg 
%\vspace*{1 cm}
\bc
\epsfig{file=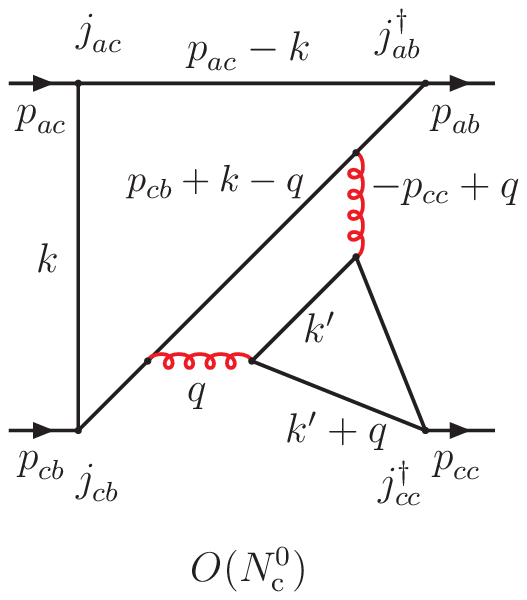,scale=0.75}
\caption{A diagram of the recombination channel of the cryptoexotic 
case with two gluon exchanges.}
\lb{af6} 
\ec
\efg
\par
The vertical cut is taken in the right part of the diagram, crossing
the four quark propagators. The Landau equations are
\bea
\lb{ea13}
& &\lambda_a^{}((p_{ac}^{}-k)^2-m_a^2)=0,\ \ \ \ \ \      
\lambda_b^{}((p_{cb}^{}+k-q)^2-m_b^2)=0,\nonumber \\
& &\lambda_c^{}(k^{\prime 2}-m_c^2)=0,\ \ \ \ \ \  
\lambda_c^{\prime}((k'+q)^2-m_c^2)=0,\\
\lb{ea14}
& &-\lambda_a^{}(p_{ac}^{}-k)+\lambda_b^{}(p_{cb}^{}+k-q)=0,\ \ \ \ \ \ 
-\lambda_b^{}(p_{cb}^{}+k-q)+\lambda_c^{\prime}(k'+q)=0,\nonumber \\ 
& &\lambda_c^{}k'+\lambda_c^{\prime}(k'+q)=0.
\eea
\par
The above system of equations can be solved, leading to the physical
singularity at $P^2=s=(m_a^{}+m_b^{}+2m_c^{})^2$, which is the signal
of the presence of four-quark intermediate states.
\par
As a last example, we consider the direct channel I diagram with two gluon 
exchanges (Figs. \rf{2f1}(b) and \rf{af7}).
\bfg 
%\vspace*{1 cm}
\bc
\epsfig{file=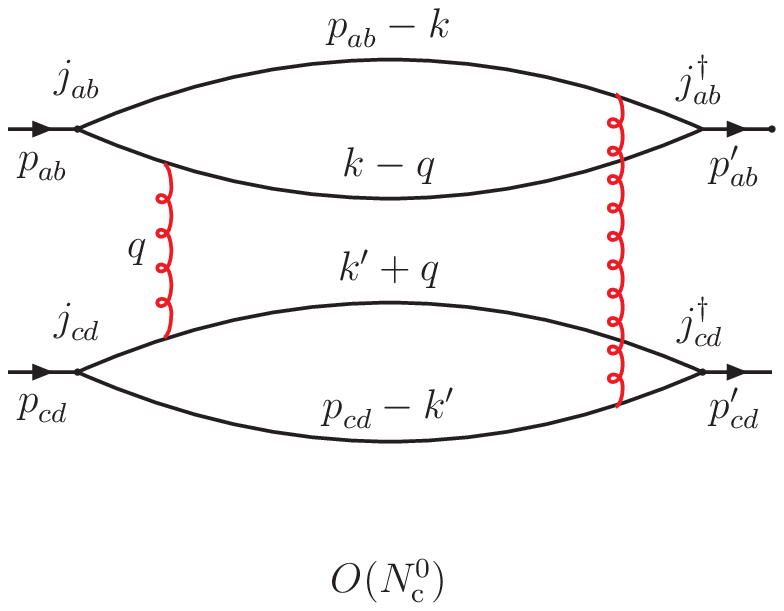,scale=0.75}
\caption{Two-gluon exchanges between the disconnected pieces of
the direct channel I [Eq. (\rf{2e3}) and Fig. \rf{2f1}(b)].}
\lb{af7} 
\ec
\efg
\par
The loop momenta can be chosen in such a way that the quark
propagators cut by the vertical line between the two gluons have the
same momenta as in the case of Fig. \rf{af5}. Therefore, one 
finds the same $s$-channel singularities.
\par
%\newpage

\end{document}